\documentclass[useAMS,usenatbib]{mn2e} 
\usepackage{deluxetable} 

\usepackage{epsfig} 
\usepackage{longtable}

\title[Radio Galaxy Evolution] 
      {Testing Models of the Individual and Cosmological Evolutions of Powerful Radio Galaxies} 

\author[P. Barai \& P. J. Wiita]
       {Paramita Barai$^{1}$ 
       \thanks{E-mail: barai@chara.gsu.edu (PB); wiita@chara.gsu.edu (PJW)} 
       and Paul J. Wiita$^{1}$ \\ 
       $^{1}$ Department of Physics \& Astronomy, Georgia State University, 
       P.\ O.\ Box 4106, Atlanta, GA 30302-4106, USA} 

\begin{document} 

\date{Accepted xxxx xxx xx. Received xxxx xxx xx; in original form 2005 Dec xx} 

\pagerange{\pageref{firstpage}--\pageref{lastpage}} \pubyear{xxxx}

\maketitle

\label{firstpage}

\begin{abstract} 
We seek to develop an essentially analytical model for the evolution 
of Fanaroff-Riley Class II radio galaxies as they age individually and 
as their numbers vary with cosmological epoch. 
Such modeling is required in order to probe in more detail the impact of 
radio galaxies on the growth of structures in the universe, 
which appears  likely to have been quite significant at $z > 1$. 
In this first paper of a series we compare three rather sophisticated analytical models 
for the evolution of linear size and lobe power of FR II radio galaxies, 
those of  \citet*{KDA}, \citet*{BRW} and \citet{MK}. 
We perform multi-dimensional Monte Carlo simulations 
in order to compare the predictions of each model for 
radio powers, sizes, redshifts and spectral indices with data. 
The observational samples used here are the low frequency radio 
surveys, 3CRR, 6CE and 7CRS,  which are 
flux-limited and complete. 
We search for and describe the best parameters for each model, 
after doing statistical tests on them. 
We find that no existing model can give acceptable fits to 
all the properties of the surveys considered, 
although the \citet{KDA} model gives better overall results 
than do the \citet{MK} or \citet{BRW} models 
for most of the tests we performed. 
We suggest ways in which these models may be improved. 
\end{abstract} 

\begin{keywords} 
galaxies: active --  galaxies: evolution -- intergalactic medium -- 
large-scale structure of universe -- methods: statistical -- 
radio continuum: galaxies. 
\end{keywords}

\section{Introduction} 

Radio galaxies (RGs) with extended lobes on opposite sides of their nuclei, 
or the classical double sources, 
constitute a significant population of active galaxies. 
\citet{FR74} classified these objects as Class II 
(FR II) sources. 
These are the edge-brightened population of radio sources 
(typically with bright hotspots), 
and also the more powerful ones, with luminosities 
$P_{178 MHz} > 10^{25}$ W Hz$^{-1}$ sr$^{-1}$. 


Flux limited samples indicate that the comoving densities of RGs were 
higher during the {\it quasar era} (i.e., between redshifts $\simeq$ 1.5 and 3) 
as compared to the present epoch \citep*[e.g.,][]{jackson99, willott01, grimes04}. 
Optical and hard x-ray observations of powerful active galactic nuclei 
reveal a similar trend for the quasar era \citep[e.g.,][]{ueda03}.
The star and galaxy formation rate was also considerably higher in the 
quasar era. 
\citet{lilly96} inferred that the observed luminosity density
(and hence the star formation rate) of the universe 
in the UV, optical and near-infrared 
increases markedly with redshift over $0 < z < 1$. 
Similarly, from Hubble Deep Field studies \citet*{connolly97} and \citet*{madau98} 
found a sharp rise in the comoving luminosity density and 
global star formation rate with redshift, 
finding that it peaked at $z \simeq 1.5$, and decreased monotonically
at higher $z$ out to $z \simeq 3 - 4$. 
More recently, \citet{bouwens06} found an apparent decrease in the 
rest-frame UV luminosity function and the cosmic star formation rate density 
from the peak redshift of $z \sim 3$ upto $z \sim 6 - 10$. 
Studies made with the Spitzer Space Telescope \citep[e.g.,][]{perez05}  
also indicate that the infrared luminosity function 
and the cosmic star formation rate increase with redshift until the quasar era. 

Submillimeter surveys claimed that the comoving luminosity density
has a peak at $z \sim 2-5$ \citep{blain99, archibald01}.
This redshift range is somewhat higher than what optical surveys
(possibly affected by dust obscuration) infer. 
At the same time, 
a more recent sub-mm study \citep{rawlings04a} indicates 
no compelling evidence for the far infrared luminosity 
of radio sources to rise with redshift. 

The above observations have prompted investigations 
of the effect of RGs on cosmological evolution and distribution of 
large scale structures in the universe. 
Preliminary work indicates that RGs can have substantial 
impacts on the formation, distribution and evolution 
of galaxies and large scale structures of the universe 
(e.g., \citealt{GKW01}, hereafter GKW01; 
\citealt{kronberg01}; \citealt{GKW03}, hereafter GKW03; 
\citealt*{GKWO}, hereafter GKWO; \citealt*{GKWB04}, hereafter GKWB; 
\citealt{rawlings04, levine05}). 


One important aspect of this process is the role played by the huge 
expanding RG lobes in triggering extensive star formation in a 
multi-phase intergalactic medium. This idea has been discussed by 
several authors in order to explain the alignment between large scale 
optical emission and radio source direction 
\citep[e.g.,][]{begelman89, deYoung89}. 
\citet{chokshi97} proposed that RG expansion could trigger much star 
formation in host galaxies. 
GKW01 stressed that RGs could impact a large fraction of the 
filamentary structures in which galaxies form, thus hastening their birth. 
Similar conclusions were drawn from different lines of argument by 
\citet{kronberg01} and \citet{furlanetto01}. 
Recently, \citet{silk05} also argued that efficient ultraluminous 
starbursts can be triggered by AGN jets. 

A very significant fraction of the volume of the universe 
in which star formation has occurred was impinged upon 
by the growing radio lobes during the quasar era 
(GKW01 and references therein). 
When these radio lobes propagating through the protogalactic medium 
envelop cooler clumps of gas embedded 
within the hotter gas which fills most of the volume, 
the initial bow shock compression triggers large-scale star 
formation, which is sustained by the persistent overpressure 
from the engulfing radio cocoon. 
This cocoon pressure is likely to be well above the equipartition 
estimate \citep{blundell00}. 
This scenario is supported by many computations, analytical 
\citep[e.g.,][]{rees89, daly90}, 
hydrodynamical 
\citep*[e.g.,][]{deYoung89, mellema02, fragile04, saxton05}, 
and magnetohydrodynamical 
\citep[e.g.,][]{fragile05}. 
This triggered star formation provides an explanation for much of 
the remarkable radio-optical 
alignment effect exhibited by high-$z$ radio galaxies 
\citep*[e.g.,][]{mcCarthy87, chambers88b}. 
Additional support for jet or lobe-induced star formation comes from 
the {\it Hubble Space Telescope} images of $z \sim 1$ radio galaxies 
\citep*{best96} and of some radio sources at higher $z$ 
\citep[e.g.,][]{bicknell00}. 
Keck observations \citep{dey97} and sub-mm observations \citep*{greve05} 
of high $z$ RGs also give evidence for this phenomenon. 
Clustered Lyman $\alpha$ emitters have been found at high redshifts 
$(z \sim 2-5)$ close to RGs \citep{venemans05, roderik05}, 
indicating that RGs form in high density regions 
and could have significant impact by accelerating star formation. 
Deep optical HST imaging gives evidence of star formation 
and a starburst driven superwind induced by AGN jet activity 
in a $z=4.1$ RG \citep{zirm05}. 

The expanding RG lobes also could have infused magnetic fields of 
significant strengths ($\sim 10^{-8}$ Gauss, e.g., \citealt*{ryu98}) 
into the cosmic web portion of the IGM 
(GKW01; \citealt{kronberg01}; GKWO; GKWB). 
Evidence of substantial metalicity in underdense regions of the IGM 
at $z \sim 4$ \citep[e.g.,][]{schaye03} 
requires a strong mechanism of spreading metals widely (metalization) 
at early cosmic epochs. 
The huge radio lobes could contribute substantially to spreading metals 
into the IGM, by sweeping out the metal-rich ISM of some young galaxies 
which they encounter while expanding (GKW03, GKWB). 

Ascertaining the importance of these processes 
of star formation, magnetization and metalization via RGs requires 
addressing the question of what fraction of the relevant 
volume of the universe did the radio lobes occupy during the quasar era 
(GKW01). 
The ``relevant universe'' refers to the volume containing 
most of the baryons, the majority of which exist as a filamentary structure of 
warm/hot gas, the WHIM (warm/hot intergalactic medium) with $10^5 <$ T $< 10^7$ K 
\citep[e.g.,][]{cen99, cen06, dave01}. 
For the radio lobes to have an important role in impacting star formation 
and spreading magnetic fields and metals, they need to occupy 
a significant portion of this relevant volume of baryons, 
which, however, was 
a small fraction of the total volume of the universe during most of the 
quasar epoch. 
A prerequisite for a more accurate computation of this RG impacted 
volume is a good model of the evolution of radio sources, 
both individually and as a function of $z$. 

Many analytical models have been published which characterize 
radio sources in terms of their dynamics and power evolution. 
\citet[ hereafter KA]{KA} showed that the 
cocoons can have a self-similar growth. 
Although numerical hydrodynamical studies \citep[e.g.,][]{carvalho02} 
indicate that radio source sizes grow in a more complex way 
than self-similar predictions, 
they are still reasonable approximations overall. 
The power evolutions in these models are 
dominated by adiabatic losses as the lobe expands, 
synchrotron radiation losses in the lobe magnetic field and 
inverse compton (IC) losses off the cosmic microwave background (CMB) photons. 

The three models of radio lobe power evolution with time which are considered 
in detail in this paper are those given by \citet*[ hereafter KDA]{KDA}, 
\citet*[ hereafter BRW]{BRW} and \citet[ hereafter MK]{MK}. 
The source linear size evolution in BRW and MK essentially follow the KDA 
prescription. They differ in the way the relativistic particles 
are injected from the jet to the lobe, and in treatments of loss terms 
and particle transport. 
So there are some significant differences in their predictions for 
observed powers ($P$) as functions of source size ($D$) and redshift ($z$). 

The simplest method to study the power evolution of 
RGs is to examine their radio power -- linear size, or $P$--$D$, diagram. 
These $P$--$D$ tracks have been used 
(KDA; MK; \citealt*{machalski04a, machalski04b}) 
to look for consistency between data and models. 
These papers compare model tracks with $P$--$D$ diagrams of observed radio 
sources to evaluate the qualitative success of the models. 

The innovative radio sky simulation 
prescription in BRW adds new dimensions to the observed parameter space. 
Using the RG redshift distribution estimated by BRW from the work of 
\citet{willott01} on the radio luminosity function (RLF) 
and any lobe power evolution model, 
one can get $P$, $D$, $z$, and spectral index $\alpha$ 
($P_{\nu} \propto \nu^{-\alpha}$) values for simulated model radio sources. 
The distributions of these simulated RGs can then 
be compared to observational data to test the success of the model. 
In BRW, slices through the 
[$P$, $D$, $z$, $\alpha$]-space generated by their model are 
{\it qualitatively} compared with observations for 
two data sets (3CRR and 7CRS); 
those authors claim good results, except for plots involving $\alpha$. 

However, to properly claim success for a theoretical model a 
{\it quantitative} statistical test is required; we present some in this paper. 
A quantitative comparison of 
cosmological radio source evolution model predictions with 
an observational data sample (the 3C data from \citealt{laing83}) 
has been done 
by \citet{kaiser99a}. 
They considered a progenitor FR II source population 
being born over cosmic epochs, and evolving according to assumed 
distribution functions of the model parameters of the KDA and KA models. 
Constructing simulated samples, they then compared the models' predictions 
with observations. 
They used $\chi^2$ statistics 
in the $[P - D]$ and $[P - z]$ planes to constrain the models. 
However the binning they used 
was somewhat arbitrary and the bins appear to be based on the 
concentration of sources in the observed $[P - D - z]$ planes. 

Our approach (based on 1- and 2-dimensional 
Kolmogorov-Smirnov (KS) statistics and correlation coefficients) 
may be as good as can be done since 
we are dealing with source characteristics in four dimensions 
($P$, $D$, $z$, $\alpha$) and 
over three observational surveys (3CRR, 6CE and 7CRS) 
with only a few hundred sources in total. 
We tried to perform multi-dimensional KS like tests (discussed in \S5.2.1) 
but the limited sizes of the observational samples 
precluded any useful results from being obtained. 

In \S2, we summarize the BRW simulation prescription. 
We apply this prescription to 
the KDA, BRW and MK models in \S3. 
In \S4 we discuss the observational samples 
to which we will compare the model distributions, and 
describe how our multi-dimensional Monte Carlo simulations are done. 
We perform statistical tests comparing the distributions of 
radio source parameters predicted by each model and those 
of observational samples in \S5. 
We vary the parameters of the models, aiming to find the parameters 
which give the best statistical fit for each model to 
all three surveys simultaneously. 
A discussion and conclusions are given in \S6 and \S7, respectively. 

\section {Initial Population Generation} 

We follow the prescription given in detail in BRW to generate the 
initial radio source population. 
Here we summarize the initial distributions of source ages, 
redshifts and beam powers; 
these produce the redshift, beam power and the age at which 
each model RG  will be intercepted by our light cone. 
This summary and update of the BRW prescription is necessary 
to define the model parameters. 
One key difference from BRW is that we assume a consensus cosmology, i.e., 
a flat universe with $H_0 = 71$ km s$^{-1}$ Mpc$^{-1}$, 
$\Omega_m = 0.3$ and $\Omega_\Lambda = 0.7$ 
\citep{spergel03}. 
The cosmological equations are taken from \citet*{carroll92} and \citet{peacock99}. 

From some initial high redshift, $z_{start}$, well above the peak of the RLF, 
sources are assumed to be born at an interval, $\Delta T_{start}$, which is 
short compared to their lifetimes. 
From the cosmology assumed 
the redshifts are translated to cosmic times (epochs) and vice versa 
\citep{weinberg89}. 
We use $z_{start} = 10$ and take $\Delta T_{start} = 10^6$ years, 
but the results should be insensitive to values of 
$z_{start} > 6$ and $\Delta T_{start} < 10^7$ years. 

After a source is born at a redshift $z_{birth}$, 
its active lifetime is denoted as $T_{MaxAge}$. 
A default value of $T_{MaxAge} = 5 \times 10^8$ years is taken. 
This value is used by BRW, and more recent investigations 
involving X-ray activity in AGN \citep{barger01}, 
SDSS optical studies of active galaxies \citep{miller03} 
and black hole demographics arguments \citep[e.g.,][]{marconi04} 
all support 
values of over $10^8$yr. 
In order to observe a radio galaxy when its nucleus is still 
actively feeding its jet, it must be intercepted by our light cone 
at some epoch between the cosmic time of its birth and the time 
when its beam is switched off, 
i.e., within an interval of $T_{MaxAge}$ after its birth. 
For this interception to occur the source must lie inside a certain 
cosmic volume shell, the ``Relevant Comoving Volume Element'', $V_C$ (BRW). 

For a spatially flat ($k = 0$) universe, 
if $r$ is the radial comoving coordinate, 
$V_C = 4 \pi R^3(t) \left( r_2^3 - r_1^3 \right) / 3$, 
where $R(t)$ is the scale factor of the universe at cosmic time $t$, 
and $r_1$ and $r_2$ are the inner and outer radial coordinates 
of the volume shell (at $t_{birth}$). 
The value of $V_C$ is the relevant volume at the epoch 
$z_{birth}$ (or $t_{birth}$). 
The corresponding proper volume now is 
$ V_C (z=0) = \left (1+z_{birth}\right)^3~V_C (z_{birth}).$ 

The sources are assumed to be distributed in redshift according 
to a gaussian RLF with (BRW Eq.~24) 
\begin{equation} 
 \rho(z) \propto {\rm exp}\left[- \frac{1}{2} 
                      \left( \frac{z-z_0}{\sigma_z} \right)^{2}\right], 
\end{equation} 
a distribution that peaks at $z_0$ and has standard deviation of $\sigma_z$. 
According to the RLF of \citet{willott01}, $z_0 = 2.2$, $\sigma_z = 0.6$, 
and we use these values in our simulations. 
\citet{grimes04} have given a more recent computation of the RLF 
where the values are $z_0 = 1.7, \sigma_z = 0.45$ 
(see their Table 5). 
The number of sources born at some 
cosmic time ($t$), per unit cosmic time, per unit comoving volume element 
at redshift zero is found from the relation $\rho(t)dt$ = $\rho(z)dz$. 
For a homogeneous and isotropic universe, this distribution is valid 
at all epochs throughout the space. 
At a particular redshift $z_{birth}$, the comoving volume ($V_C$) is found. 
Then multiplying $V_C$ by $\rho(z_{birth})$ gives the number of sources born 
at $z_{birth}$ (per solid angle) in the chosen interval in cosmic time 
which are intercepted by our light cone: 
$ N_{born} \propto V_C(z=0)  ~ \rho(z_{birth}).$ 
The total number, $N_{born}$, is obtained by using a normalization factor 
in the above proportionality which takes into account the sky area of 
the observed data sample. 

Homogeneity of the universe implies that the sources are randomly 
distributed within the comoving volume shell. 
The age of a source, $T_{age}$, is the time after $t_{birth}$ 
it is intercepted by our light cone; in our computations it is drawn 
randomly from $0$ to $T_{MaxAge}$, but weighted so that 
sources are distributed uniformly in volume within the comoving volume shell. 

In each simulation (run) we have generated a very large number of sources, 
over a wide range of cosmic time. 
We find the number of sources born at some $z_{birth}$ 
which will intercept our light cone, the age $T_{age}$ 
(denoted by $t$ henceforth) of each source, and 
the redshift at which we observe it (denoted by $z$ henceforth), 
which is derived from $T_{obs}$, the cosmic time at which the 
light we see was emitted from the source. 

As very powerful sources are much rarer than weaker ones, 
each of the sources generated is assigned a jet power $Q_0$ 
(which is assumed to remain constant throughout its age) according to the 
probability distribution (BRW Eq.~38)
\begin{eqnarray} 
p(Q_0)dQ_0 & \propto & Q_0^{-x}~dQ_0 ~\textrm{ if $Q_{min} < Q_0 < Q_{max}$}, \nonumber \\ 
           &    =    & 0 ~~~~~\textrm{ if $Q_0 > Q_{max}$ or $Q_0 < Q_{min}$}. 
\end{eqnarray} 
Here the power index $x$ is positive, and we initially adopted the values used by BRW: 
$x = 2.6, Q_{min} = 5 \times 10^{37}$ W, and $Q_{max} = 5 \times 10^{42}$ W. 
Our best fit values of $x$ are higher and are discussed in \S5. 

An initial Monte Carlo population generation is completed when 
$t$, $z$ and $Q_0$ are randomly assigned to each source of 
the population according to the above prescriptions. 
Each source in that population is then allowed to evolve according to a model described 
in the following section, giving the observable quantities other than $z$: 
$P$, $D$ and $\alpha$. 

\section {Models of Radio Lobe Evolution} 

A standard basic model of FR II extragalactic radio sources 
\citep[e.g.,][]{scheuer74, blandford74} is widely accepted. 
A powerful RG consists of the central active nucleus 
and two jets emerging from opposite sides of it. After traveling 
substantial distances the plasma in these jets collides with a tenuous environment. 
There the jets terminate in a 
shock 
where relativistic electrons are accelerated and hotspots are formed; 
the plasma passing through the terminal shocks inflate the huge 
lobes of energetic particles. 
A bow shock propagates into the surrounding gas ahead of the jets. 

The radio power evolution models that we compare are those given by KDA, BRW and MK. 
In brief, the physics of these models differ 
mainly in the ways in which particles are assumed to be transported from the jet 
through the hotspot and into the lobe. 
KDA assume a constant injection index, $p$, for the energy number distribution 
so $N(E) \propto E^{-p}$, 
for the radiating relativistic particles 
while the particles are injected from the hotspots into the lobes. 
BRW assume that the injection index varies between the different 
energy regimes, as governed by the break frequencies discussed below. 
MK assume a constant injection index but also argue that the particles are 
re-accelerated by some turbulent process in the head during transport 
to the lobes. 
Several key points of each model and additional differences are 
noted in \S\S 3.2 -- 3.4, 
although the reader should refer to the original papers for a thorough 
understanding of each model's details. 
Table 1 lists the default values of the major model parameters 
(those used by the authors). 
We varied these parameters 
in our extensive simulations described in \S4.3. 
The only parameter whose variation was not considered is the adiabatic index 
of the external environment, which was adopted as usual as $\Gamma_x=5/3$. 


\begin{table}
\caption{Default Values of the Model Parameters\label{tab1}\tablenotemark{a}}
\begin{tabular}{cccc}
\hline 
Parameter              &  KDA      &  BRW      &  MK \\ 
\hline 
$\beta$                & 1.9       & 1.5       & 1.5 \\
$a_0$ (kpc)            & 2         & 10        & 10 \\
$\rho_0$ (kg m$^{-3}$) & $7.2\times10^{-22}$ & $1.67\times10^{-23}$ & $1.7\times10^{-23}$ \\
$\Gamma_x$             & 5/3       & 5/3       & 5/3 \\
$\Gamma_c$             & 4/3       & 4/3       & \\
$\Gamma_B$             & 4/3       &           & \\
$R_T$                  & 1.3       &           & \\
$\gamma_{min(hs)}$     & 1         & 1         & 10 \\
$\gamma_{max(hs)}$     & Infinity  & $10^{14}$ & $10^7$ \\
$p$                    & 2.14      & 2.14      & 2.23 \\
$r_{hs}$ (kpc)         &           & 2.5       & 2.5 \\
$t_{bs}$ (yr)          &           & $10^5$    & \\
$t_{bf}$ (yr)          &           & 1         & \\
$\eta$                 &           &           & 0.4 \\
$\epsilon$             &           &           & 1.0 \\
$\tau$                 &           &           & $2\times10^{-3}$ \\
\hline
\tablenotetext{a}{See text and the original papers for parameter definitions.}
\end{tabular}
\end{table}


\subsection {Dynamical Expansion and Emission} 

In all of the models we consider here the ambient gas 
around the double radio sources, into which the lobes 
propagate, is taken to have a power-law 
radial density distribution scaling with distance $r > a_0$ from 
the center of the host galaxy (Eq.~2 of \citealt{KA}), 
\begin{equation} 
\rho(r) = \rho_0 \left( \frac{r}{a_0} \right) ^ {-\beta} 
\end{equation} 
where the central density, $\rho_0$, scale length, $a_0$, and 
radial density index, $\beta$, are given by the particular model. 
We follow BRW and assume that the external 
density profile is invariant with redshift. 
While such a typical radial density distribution is 
appropriate on average for small $z$, 
this may not to be a good approximation at 
the redshifts corresponding to the quasar era, which 
witnessed a $10^2$--$10^3$ times higher co-moving density of powerful radio-loud 
ellipticals \citep[e.g.,][]{jackson99}. 
We note that for very large sources the density will depart from a single 
power-law with radius and eventually approach a 
constant value appropriate to the intergalactic medium at that redshift 
(e.g., \citealt{GKW87}; \citealt{furlanetto01}). 
If the ambient density approaches a constant value 
at radial length scales of $\sim$ 100 kpc, 
then radio sources grow to somewhat smaller sizes and have larger lobe powers. 
We will consider this more complicated situation in future work.

From dimensional arguments 
(\citealt[][ or KA]{KA}; \citealt{komissarov98}) 
the total linear size (from one hotspot to the other) of a radio source 
at an age $t$ can be expressed as 
\begin{equation} 
D(t) = 2 c_1 a_0 \left( \frac {t^3 Q_0} {a_0^5 ~ \rho_0} \right)^{1/(5-\beta)}, 
\end{equation} 
where $c_1 \sim 1$, is model dependent, but weakly varying, as discussed below. 
The jump conditions at the external bow shock and the expression for linear size 
give the pressure of the head plasma immediately downstream of the bow shock as (KA Eq.~12) 
\begin{equation} 
p_h(t) = \frac {18 c_1^{2-\beta}} {\left(\Gamma_x +1\right) \left(5-\beta\right)^2} 
\left( \frac { \rho_0^3 a_0^{3\beta} Q_0^{2-\beta} } {t^{4+\beta}} \right)^{1/(5-\beta)}. 
\end{equation} 

An ensemble of $n(\gamma)$ relativistic electrons with Lorentz 
factor $\gamma$ 
in a volume $V$ with magnetic field $B$ emits 
synchrotron power per unit frequency, per unit solid angle given by 
(KDA Eq.~2) 
\begin{equation} 
P_{\nu} = \frac{\sigma_T c}{6 \pi} \frac{B^2}{2 \mu_0} 
          \frac{\gamma^3}{\nu} n\left(\gamma\right) V 
\end{equation} 
in units of W Hz$^{-1}$ sr$^{-1}$, with $\sigma_T$ the Thomson 
cross-section and $\mu_0$ the permeability of free space. 
These relativistic electrons are injected into the lobe 
from the hotspot via the head, 
an extended region of turbulent acceleration around the hotspot. 

\subsection {The KDA Model} 

For the density profile of the external atmosphere this model uses 
$\rho_0 = 7.2 \times 10^{-22}$ kg m $^{-3}$, $a_0 = 2$ kpc and 
$\beta = 1.9$. These values are argued to be typical for an elliptical 
galaxy out to $\approx 100$ kpc from its center \citep*{forman85, canizares87}. 
The factor $c_1$ (Eq.~4) 
is given by Eq.\ (32) of KA, which by their 
Eqs.\ (37) and (38) depends weakly on $R_T$, the axial ratio, 
defined as the ratio of the length of the source and its width. 
For $R_T=1.3$, the value adopted by the authors and us, $c_1=1.23$. 

The ratio of pressure in the head to that in the cocoon was 
taken by KDA to be (Eq.~38 of KA) 
$p_h / p_c = 4 R_T ^ 2 $. 
We follow this prescription; 
however, the hydrodynamical simulations of \citet{kaiser99b} 
found this ratio to be an overestimate. 
The ``improved'' KDA model \citep{kaiser00} obtains an 
empirical formula for this ratio as (Eq.~7 of \citealt{kaiser00})
$p_h / p_c = \left(2.14 - 0.52 \beta \right) R_T^{2.04-0.25\beta}$. 
We are exploring this alternative approach and will give results using it in 
the next paper in this series. 

The electrons are assumed to be accelerated in the hotspot at 
time $t_i$, with corresponding initial Lorentz factor $\gamma_i$. 
The energy distribution of the electrons injected into the lobe is a 
power law function of $\gamma_i$, $n(\gamma_i) \propto \gamma_i ^ {-p}$; 
$p$ is taken to be constant. 
The electron energies evolve in time according to (Eq.~4 of KDA) 
\begin{equation}
\frac{d\gamma}{dt} = -\frac{a_1 \gamma}{3t} - 
                      \frac{4 \sigma_T}{3 m_e c} \gamma^2 \left(u_B + u_c\right).
\end{equation} 
Here the lobe electrons undergo energy losses via 
adiabatic expansion 
($V \propto t^{a_1}$ with 
$a_1 = \left( 4+\beta \right) / \left[ \Gamma_c \left( 5-\beta \right) \right]$ 
and $\Gamma_c$ the adiabatic index in the cocoon, KDA), 
IC scattering off the CMB photons and synchrotron losses. 
The magnetic field (assumed to be completely tangled) with energy density $u_B$ 
and adiabatic index $\Gamma_B = 4/3$, satisfies 
$ u_B \propto B^2(t) \propto t^{- \Gamma_B a_1} $. 
The energy density of the CMB, $u_c$, is taken to be constant for an individual 
radio source as each source evolves 
for only a few times $10^8$ years. 


The KDA model does not distinguish between the head and hotspot, 
and considers self-similar expansion of the head, where the jet terminates. 
The cocoon is split into many small volume elements, each of which is 
allowed to evolve by expanding adiabatically 
(changing the pressure from head pressure $p_h(t_i)$ to cocoon pressure $p_c(t_i)$) 
and undergoing the various loss processes. 
The energy of each volume element in the lobe is equated to the 
energy it had while in the head minus the work done by the volume in 
adiabatically expanding from the head to the lobe. 
The radio emission from such a volume element is calculated, 
using the expressions of cocoon pressure and the energy distribution function. 
The total emission at a frequency $\nu$ is then obtained by summing over the 
contributions from all such small elements in the lobe. 
The expression for $P_{\nu}$, given by Eq.\ (16) of KDA is a complicated 
integration over injection time $t_i$. 
This integration being analytically intractable, 
we used numerical techniques to get $P_{\nu}$. 

\subsection {The BRW model} 

The ambient gas density parameters adopted by BRW are 
$\rho_0 = 1.67 \times 10^{-23}$ kg m$^{-3}$, $a_0 = 10$ kpc and 
$\beta = 1.5$. 
These are based on polarization measurements of lobe 
synchrotron emission \citep{garrington91}, 
and X-ray images of massive ellipticals \citep[e.g.,][]{sarazin88, mulchaey98}. 
A value of $c_1 = 1.8$ is adopted 
in Eqs.~(4) and (5), as BRW found it to give the best fit between models and data. 

This model assumes the hotspot to be a compact region 
(the working surface moving around as in 
Scheuer's \citeyearpar{scheuer82} ``dentist's drill model'') 
within the whole head region. Considering the expansion of the head and 
its bow shock \citep[also][]{begelman89}, the environmental ram 
pressure is related to the average internal pressure in the head (Eq.~5). 
The pressure in the lobe in taken to be a constant factor (1/6) 
of the head pressure. 

The jet, of constant bulk power $Q_0$, terminates at the hotspot, 
which is taken to be of constant radius, $r_{hs} = 2.5$ kpc, in BRW. 
The pressure in the hotspot, $p_{hs}$, is given by the stagnation 
pressure in the post-jet shock, $p_{hs} = Q_0 / (c A_{hs})$. 
Here $A_{hs}$ ($= \pi r_{hs}^2$) is the area normal to the jet over 
which the jet thrust operates. 
The hotspot magnetic field, assumed to be tangled, is given by 
$B_{hs}^2 = 3 \mu_0 Q_0 / (c A_{hs})$, 
where the equipartition assumption has been made. 
The break frequency for synchrotron radiation in the hotspot is 
(Eq.~12 of BRW) 
\begin{equation} 
\nu_{bh} = \frac {9 c_7 B_{hs}} {4 \left( B_{hs}^2 + B_{CMB}^2 \right)^2 t_s^2}, 
\end{equation} 
where $c_7$ is $1.12 \times 10^3$ nT$^3$ Myr$^2$ GHz \citep{leahy91}, 
and the equivalent magnetic field due to the CMB is 
$B_{CMB} = 0.318 (1+z)^2$ nT. 
The synchrotron age, $t_s$, of the electron population is determined by 
the length of their exposure to the hotspot magnetic field 
before they reach the lobe. 
This longest dwell time in the hotspot is taken as $t_{bs} = 10^5$ yr, 
and the shortest dwell time $t_{bf} = 1$ yr.

In \S8.4.2 of BRW it is shown that this model roughly follows the 
KDA prescription of lobe luminosity, but with two main differences. 
First, while the particles are injected from the hotspot to the lobe, 
the injection index is governed by the breaks in the energy distribution of 
particles (unlike the constant injection index of KDA). 
Second, the constant hotspot pressure governs the 
adiabatic expansion losses out of the hotspot 
(for particles injected into the lobe), 
while in KDA the head pressure (which evolves with time) 
drove the adiabatic losses. 
In BRW the head pressure (Eq.~5) only drives the source expansion. 

The details of the energy distribution 
(as a function of the  Lorentz factor) of particles in the hotspot 
are shown in Fig.\ 11 of BRW. 
Our calculations usually are done assuming the minimum and maximum values of 
the particle Lorentz factors in the hotspot quoted by BRW: 
$\gamma_{min(hs)} = 1$ and $\gamma_{max(hs)} = 10^{14}$, 
although we have examined variations in these parameters. 

A population in the lobe which emits at a time $T_{obs}$ 
(when it intercepts our light cone), consists of 
particles injected from the hotspot between a time $t_{min}$ 
(those with largest Lorentz factors) and $T_{obs}$ (smallest Lorentz factors). 
The time $t_{min}$ (found following the prescription in KDA) is the 
minimum time of injection (found for every $T_{obs}$), 
when particles can still contribute to the radiation at $\nu$. 

The final expression for the power emitted ($P_{\nu}$) by a radio source 
at a frequency $\nu$ is given by the complicated Eq.\ (21) of BRW, 
which we will not reproduce here. We solved this equation numerically. 

\subsection {The MK model} 

The \citet{MK} paper employs the same external density profile and 
source linear size expansion as does BRW. 

The MK model essentially follows the common prescriptions of KDA and BRW for 
lobe luminosity evolution, with the key difference involving  the 
particle transport mechanism. 
Two cases are considered for the propagation of particles from the 
termination shock through the hotspot and into the cocoon. 
In MK's Case A, the whole adiabatic loss between the hotspot and lobe 
(due to the pressure difference) is computed. 
However, the authors found that this produced $P - D$ tracks which 
conflicted with the observational data. 
So they considered a Case B, which involves some re-acceleration process 
in the turbulent head region, whereby the adiabatic losses are partially compensated; 
MK found such a model is a qualitatively better fit to the data. 
Thus we consider only the case B (with re-acceleration) of the MK model in our present work. 

This model assumes that electrons are accelerated by the first-order Fermi mechanism 
at the jet termination shock and are injected into the plasma behind the shock 
following a power-law energy distribution with a constant injection index $p$. 
A fraction $\eta$ of the jet power is assumed to be transferred 
into the accelerated particles at the termination shock. 
If the bulk Lorentz factor of the jet, $\gamma_{jet} \sim 10$, 
then $ 2 < p < 2.3 $ \citep[e.g.,][]{achterberg01}. 
The correct upper and lower limits of particle Lorentz factors 
$\gamma_{min}$ and $\gamma_{max}$ are not obvious; 
MK adopt $\gamma_{min} = \gamma_{jet}$. 
The authors say the results are not sensitive to $\gamma_{max}$; 
however, our different conclusions on this point are given in \S\S5-6. 

After being dumped in the primary hotspot by the jet, the electrons 
encounter turbulent motions of the plasma in transit through the head 
and finally reach the lobe. 
In this transition through the head the electrons are subject to 
synchrotron losses (in the strong magnetic field behind the termination shock) 
and IC losses off the CMB. 
The effects of the losses depend on the distribution of the ``escape times'', i.e., 
the probability distribution of how many particles escape after a certain 
time interval. 
A generalized transport process is considered, with $\epsilon$ 
(denoted as $\alpha$ in MK) being the transport parameter (or the diffusion index). 
The mean square distance traveled by a particle, 
$\langle \Delta r^2 \rangle \propto t^{\epsilon}$, with $0 < \epsilon < 2$. 
In the standard diffusive case, $\epsilon = 1$, with sub- (supra-) 
diffusive cases being $\epsilon < 1$ ($> 1$). 
Another new parameter in this model is $\tau$, 
the ratio of the diffusive transport time and cooling time 
of a particle at $\gamma_{min}$.  

During the transport of particles from hotspots to lobes, the details of re-acceleration 
by various processes have been considered by many previous authors 
\citep*[e.g.][]{spruit88, begelman90, manolakou99, gieseler00}. 
MK simply assume that in the presence of reacceleration, 
the distribution of electrons entering the lobe is described by a power law 
above a lower cut-off energy, and at higher energies is modified by 
synchrotron and IC losses. 

Once the electrons have reached the lobe, they undergo adiabatic, 
IC and synchrotron losses, similarly to the other models, 
and their energy evolution is given by Eq.~22 of MK. 
Similarly to the KDA model, for every time instant $t$, when radiating particles have 
Lorentz factor $\gamma$, there is an earliest time, $t_i$, at which 
particles injected into the lobe contribute to the radiation at $t$; 
$t_i$ can be obtained following the prescription given in Eqs.\ (25) and (26) of MK. 

The final expression for power emitted at a frequency $\nu$, $P_{\nu}$ 
is given in Eq.\ (27) of MK. 
The authors used $r_{hs} = 2.5$ kpc as the hotspot radius; 
however, we find that the MK model power results are actually 
independent of the hotspot area $A_{hs}$. 
From Eq.\ (2) in MK, $u_h \propto 1/A_{hs}$ and from 
their Eq.\ (6), $t_0 \propto A_{hs}^{1/a}$, 
where $a = \left(4 + \beta\right) / \left(5 - \beta\right)$. 
Hence $u_{lobe}$ (MK Eq.\ 5), $b_s$ (MK Eq.\ 22) and 
$P_{\nu}$ are independent of $A_{hs}$. 

\section {Observational and Simulated Samples} 

\subsection {Selection Criteria and Observed Characteristics} 

These models predict the emission from the radio lobes, 
which are taken to (and usually do) 
dominate the emission from extended FR II RGs. 
As is well known and is discussed in detail in BRW, 
at relatively low frequencies ($\sim 151$ MHz) 
the radio flux observed is predominantly the emission from the cocoon or the lobe 
(with negligible contribution from the hotspots, jets or nucleus), and so these 
evolutionary models should fit the data best at such frequencies. 
At GHz frequencies, substantial contributions from Doppler boosted core or jet 
emission would often be present, especially for old quasars, 
but the slowly advancing lobes will still emit nearly isotropically. 
In addition, at these higher frequencies the effects of synchrotron, adiabatic and 
IC losses are more severe. 
At very low frequencies ($< 100$ MHz), there are extra complications affecting 
the emission from synchrotron self-absorption, free-free absorption, 
and the poorly known low energy cut-off to the relativistic synchrotron 
emitting particles. 
Therefore samples such as those produced by the Cambridge group over the 
past decades, which were observed at between 151 and 178 MHz, 
and cover much of the northern sky, are most appropriate for this work. 

We adopt observational samples from complete radio surveys (Table 2), 
each of which contains all 
the radio sources within each survey's flux limits and which are found inside 
smaller sky areas, for deeper surveys. 
Redshifts have been obtained 
for the great majority of these radio sources. 
This lower flux limit brings in a $P - z$ correlation, 
since $P$ decreases as $z$ increases. 
To decouple this $P - z$ correlation one must use multiple complete samples 
at increasingly fainter flux limits. 


For an individual source in each survey, the following characteristics were 
considered: 
the redshift ($z$), the specific power at $151$ MHz 
($P_{151}$) in W Hz$^{-1}$ sr$^{-1}$, 
the total projected linear size ($D$) in kpc, 
and the spectral index at $151$ MHz ($\alpha_{151}$) converted to the rest frame of the source. 
The redshifts in the samples are spectroscopically determined for 
the vast majority of the sources. 
For the 3CRR catalog the redshift completeness is $100 \%$, 
for 6CE it is $98 \%$ and it is $92 \%$ for 7CRS. 

\subsection {Observational Sample Details} 

Henceforth, 3C, 6C, and 7C refer to the refined surveys 3CRR, 6CE and 7CRS, 
respectively, as described below. 
We excluded FR I RGs from the following catalogs and 
considered only FR II sources (including quasars, weak quasars, 
high-excitation FR II RGs and low-excitation FR II RGs). 


\begin{table}
\caption{Observational Samples\label{tab2}}
\begin{tabular}{cccccccc}
\hline
Survey & Flux Limit & No. of Sources\tablenotemark{a} &  Sky Area \\
       & (Jy)       &                                 & (sr)      \\
\hline
3CRR  & $S_{178} \tablenotemark{b} ~ > 10.9 $ & 145 & 4.23 \\
      & $S_{151} > 12.4 $ \\
\\
6CE   & $2 \leq S_{151} \leq 3.93 $ & 56 & 0.102 \\
\\
7CRS  & $S_{151} > 0.5  $    & 126 & 0.022 \\
7CI   & $S_{151} \geq 0.51 $ & 37  & 0.0061 \\
7CII  & $S_{151} \geq 0.48 $ & 37  & 0.0069 \\
7CIII & $S_{151} > 0.5  $    & 52  & 0.009  \\
\hline
\tablenotetext{a}{Only FR II RGs considered.}
\tablenotetext{b}{Flux at 178 MHz, the frequency at which the 3CRR survey was performed. $S_{178}$ for these sources are converted to flux at 151 MHz, $S_{151}$, using a constant spectral index of 0.8.}
\end{tabular}
\end{table} 


3CRR: 
This is the Third Cambridge Revised Revised sample of extragalactic 
radio sources \citep*{laing83}. 
We adopted the data 
from the online compilation of the list by 
Willott\footnote{http://www-astro.physics.ox.ac.uk/$\sim$cjw/3crr/3crr.html}. 
In 3CRR the observations were done at a frequency of $178$ MHz, so for 
each 3CRR source $P_{178}$ (specific power at $178$ MHz) 
was obtained and then converted to $P_{151}$ 
using a standard average spectral index of $0.8$. 
Given the closeness of these two frequencies, any reasonable variations in $\alpha$ would 
make for only small differences in the derived $P_{151}$ values. 

6CE: 
The Sixth Cambridge radio survey by \citet{eales85} 
is the original 6C survey. 
We adopt the sample from the reselected and updated version in 
\citet*{rawlings01}, along with the most recent redshifts, 
which have been updated online by 
Rawlings\footnote{http://www-astro.physics.ox.ac.uk/$\sim$sr/6ce.html}. 


7CRS: 
The Seventh Cambridge Redshift Survey is a combination of parts I, II and III 
of the original 7C survey \citep{mcGilchrist90}. 
For 7C-I and II we adopt $P_{151}$ and $z$ from \citet{willott03}; 
(their Tables 2 and 3, which use the present consensus cosmology). 
The values of $D$ 
were 
obtained from a web-site maintained by Steve 
Rawlings\footnote{http://www-astro.physics.ox.ac.uk/$\sim$sr/grimestable.ascii}; 
however, the $\alpha_{151}$ values 
are not available in a collated form in the literature, 
and only a few individual sources have these values published. 
Thus we used $\alpha_{151}$ for 7C-III only. 
For 7C-III, the reduced data, including redshift, flux density in Jy, 
angular size in arcsec and spectral index between 38 and 151 MHz were kindly 
provided to us by Chris Willott; 
from these we computed the relevant observational parameters in the cosmology we use. 
The observed spectral index between 38 and 151 MHz was taken as the rest-frame 151 MHz 
spectral index (a fairly good estimate, at least for the higher $z$ sources). 
The relevant sample can be found in Table 9 
(containing both the 7CIII and NEC samples) of \citet{lacy99} 
or online from the website of Oxford 
University\footnote{http://www-astro.physics.ox.ac.uk/$\sim$cjw/7crs/7crs.html} (but with a different cosmology). 



\subsection {The Simulated Surveys} 

Large radio source populations are randomly generated, according to the 
source age, redshift and beam power distributions as given in \S2, 
for each choice of model parameters. 
Each simulated source, in a population, 
is then allowed to evolve in age according to a 
power evolution model discussed in \S3. 
The evolution is to be done at a rest frame frequency of the source, 
so if the frequency of observation is $\nu_{obs} = 151$ MHz, 
and a source is observed at redshift $z$, then it is evolved at 
a frequency $\nu_{rest} = 151 \times (1+z)$ MHz. 

The monochromatic power ($P_{151}$ in W Hz$^{-1}$ sr$^{-1}$) each 
source would emit at the $T_{obs}$ corresponding to it is calculated; 
this depends on the model as described in \S3. 
At this cosmic time ($T_{obs}$), its redshift, and hence its distance 
from us, is found. 
The flux (in units of Jy = $10^{-26}$ W Hz$^{-1}$ m$^{-2}$) of this 
source is then obtained 
(using \citet{peacock99}: Eqs.\ 3.87, 3.76 and 3.10 for a flat universe), 
given that it emitted $P_{151}$ from the cosmic distance calculated. 
If the flux for a source is greater than a (lower) survey flux limit (or between 
two flux limits in the case of 6C) then 
that source is considered to be detected in the corresponding simulated survey, 
and counted for the later comparisons with real data. 

It is assumed in our simulations that the radio jets 
feeding the lobes (or the central AGN) stay ``on'' 
only for the time $T_{age}$ corresponding to each source (\S2), 
which is also taken as the lifetime of a source. 
After the time $T_{age}$, the relativistic plasma in the lobes 
continue to radiate but 
the flux drops very rapidly 
once the central engine has stopped to feed the lobes. 
So the sources can be considered to be turned ``off'' instantaneously after $T_{age}$. 
This assumption is supported by the fact that the 
radio powers ($P_{151}$) drop substantially while the jets 
are still on (i.e., within the time $T_{age}$ after birth), 
as shown by the $P - D$ tracks in \S5.1. 

To perform our simulations we initially generate 
an ensemble of a huge number (a few $10^6$) of pseudo-radio sources. 
After evolving each source by the above prescription, 
the ensemble is examined to see how many of them would actually be 
detected in a simulated complete survey. 
The population size is then chosen for this parameter set in the next run 
so as to get a 
comparable number detected in the simulations as are found in the real surveys. 
To do this, we usually had to generate such ``standard'' ensembles containing 
$\approx 10^6$ to $10^7$ radio sources. 
Assuming the observed regions are fair samples of the universe, 
the population size is proportional to the sky area of a survey. 
The populations needed in order to simulate the 6C and 7C surveys 
are generated from that of 3C by reducing the total 3C population size 
according to the corresponding sky area ratio. 
Given a 3C initial ensemble of size $S_{3C}$, 
the populations for 6C and 7C are created by plucking sources 
randomly from that initial ensemble, and producing populations 
of sizes $S_{6C} = S_{3C} / 41.5$ and $S_{7C} = S_{3C} / 192.3$. 

The initial populations generated for comparison with the 6C and 7C data 
following the above procedure detected more or less comparable numbers 
of sources in the simulations as compared to the actual surveys. 
We compute the over (-under) detection factors, defined as 
the ratios of the number of sources detected in the simulated 
6C and 7C surveys to the numbers in the actual surveys 
divided by the same ratio for the 3C survey. 
The deviation of these ratios from 1.0 (see discussion in \S6) may be 
considered a measurement of the statistical (sample) variance. 


Each model gives the total linear size of a radio source, 
but we observe each one as projected on the plane of the sky. 
This is incorporated into the simulations as follows. 
Sources are considered to be randomly oriented in the sky, 
with the angle to the line of sight ($\theta$) of each source 
drawn from a distribution uniform in $(1-\cos \theta)$. 
The projected length of each simulated source is then, 
$D_{proj} = D(t) \times \sin \theta = D(t) \times \sqrt{r_N(2-r_N)}$, 
where $r_N$ is an uniform random number between 0 and 1, and 
$D(t)$ is the total linear size of the source. 
For compactness, hereafter we denote the projected size 
$D_{proj}$ as just $D$. 

The initial population of sources was generated and 
each lobe power evolution model was implemented in C, 
and the other supporting codes were written in IDL. 
Numerical Recipes in C \citep{press02} were used to speed up the 
calculations of lobe powers for the huge ensembles of sources. 

In doing the statistical tests (\S5.2 and \S5.3) 
we compared the model predictions with the observational samples as follows. 
In a single run a random initial population of millions of sources was generated 
such that 
after evolution of each source 
in the ensemble and after comparing each to the flux limits, 
the ensemble produced simulated samples (for the 3C, 6C and 7C catalogs) 
which were of sizes comparable to or larger than the real surveys. 
The simulated samples 
were then reduced in size, if necessary, by uniformly selecting sources from them. 
In particular, this was done by selecting every 
($N_{sim}/N_{samp}$)'th source from a simulated survey, 
where $N_{samp}$ is the number of sources in one of the real surveys 3C, 6C or 7C 
and $N_{sim}$ (usually $> N_{samp}$), is the number of sources in the simulated survey. 
Finally statistical tests (whose results are tabulated) 
were done on the $[P, D, z, \alpha]$ data from the real surveys 
and a similar sized simulated sample generated from a single random seed. 

Each of the $[P - D - z - \alpha]$ plane figures, described in \S6, 
show the final simulated sample 
(after reduction to the actual data sample sizes) 
of the random run done using specific parameters for each semi-analytical model. 
The plotted cases were among the best overall fits for each model 
as determined by the KS tests. 

\section {Results} 

\subsection {$P- D$ Tracks} 

\begin{figure}
\centerline{\epsfig{file=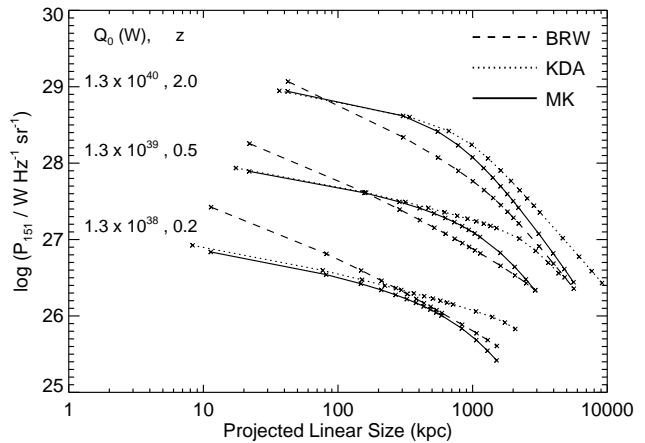, scale=0.5}}
\caption{$P - D$ tracks of three sources with jet powers ($Q_0$) in Watts
and redshifts ($z$) of $[1.3 \times 10^{40}, 2.0]$,
$[1.3 \times 10^{39}, 0.5]$, $[1.3 \times 10^{38}, 0.2]$ (from top to bottom).
Each of the {\it dashed}, {\it dotted} and {\it solid} curves correspond to
the tracks predicted by the default versions of the BRW, KDA and MK models respectively.
The crosses on the tracks denote source lifetimes of
1, 10, 20, 30, ..., 90, 100, 150, 200, 250, 300 Myr.}
\label{fig1}
\end{figure}

As a radio source gets older, its power ($P$) vs. linear-size ($D$) 
track becomes steeper. While this is true for all models, 
the rate of steepening is different in the three models, 
as seen from Fig.\ 1. 
These $P$--$D$ tracks have been generated using the default parameters 
of each model (given in Table 1), by allowing each source 
(with beam powers and redshifts given in the plot) 
to evolve at frequency $\nu=151$ MHz. 
For this Figure (alone) the total linear sizes were converted to the projected sizes 
assuming an average viewing angle to the line of sight of 
$39.5^{\circ}$ (following KDA). 
These tracks are in agreement with the conclusion drawn by MK 
that their $P$--$D$ tracks are more akin to those presented by KDA. 
Crude evaluations of the quality of different models and the allowable 
ranges of parameters for them can be found by comparing the regions in the 
$P$--$D$ diagram that are actually populated with those that are 
accessible to models with those parameters (e.g., KDA, MK). 


On examining different tracks, it is found that the luminosity 
falls off faster for sources with high power and high redshift. 
The higher the redshift of a radio source, 
the shorter the fraction of its life which can be detected by 
flux limited radio surveys. 
This point was noted by \citet*{GKW89} 
and was stressed by \citet{blundell99}, 
who coined the phrase {\it youth-redshift degeneracy} to describe it. 


\subsection{1-Dimensional Kolmogorov-Smirnov Tests} 

\subsubsection{Statistical Tests and Default Parameter Results} 


For our first attempt to quantitatively compare 
the simulated radio surveys to the actual data, 
we used 1-dimensional Kolmogorov-Smirnov (1-D KS) statistical tests. 
Based on the results of such tests we chose some parameter variations 
for the models on which two more statistical tests were done (\S5.3). 

In the 1-D KS tests, 
each of the distribution's key characteristics $[P, D, z, \alpha]$ 
of the radio sources detected in the 
simulated surveys were compared to those of the sources 
in the real radio surveys 3C, 6C, and 7C, 
according to procedures given at the end of \S4.3.
The KS probabilities, $\cal P$, that the two data sets 
being compared are drawn from the same distribution function, were taken 
to be a figure of merit of each model used in the simulation. 
High values of ${\cal P}$ (close to 1.0) indicate good fit, 
and very small ${\cal P}$ imply that the model and 
data distributions are significantly different. 
We consider twelve test statistics in total 
(the twelve probabilities found from the KS statistics for comparisons of 
each of $P, D, z, \alpha$ for each of 
the three radio surveys), 
which quantifies the closeness of the model fits to the data. 

In order to quantify the overall success of a model we would prefer to have 
a single figure-of-merit instead of twelve individual ones, but there is no 
obvious way to produce such a statistic, particularly since the three 
surveys have significantly different numbers of objects. 
A likelihood type test would involve the product 
instead of the sum of the KS probabilities, 
but given the extremely small values of these products 
we rejected this figure of merit
as not providing a useful discrimination between the models. 
Here we have combined the 1-D KS probabilities in two ways. 

First, we add the KS probability statistic for comparisons of $P, D, z, \alpha$ 
(i.e. ${\cal P}(P)+{\cal P}(D)+{\cal P}(z)+{\cal P}(\alpha)$) for the three surveys, 
weighting the statistic of a survey by the square-root 
of the number of simulated sources detected in that survey. 
So the first overall figure of merit of a model, 
which we denote as ${\cal P}_{[P, D, z, \alpha]}$, is given by: 
\begin{eqnarray} 
{\cal P}_{[P, D, z, \alpha]} =  
\left[ {\cal P}(P)+{\cal P}(D)+{\cal P}(z)+{\cal P}(\alpha) \right]_{3C} +  \nonumber \\ 
\sqrt{\frac{N_{6C}}{N_{3C}}}\left[ {\cal P}(P)+{\cal P}(D)+{\cal P}(z)+{\cal P}(\alpha) \right]_{6C} + \nonumber \\ 
\sqrt{\frac{N_{7C}}{N_{3C}}}\left[ {\cal P}(P)+{\cal P}(D)+{\cal P}(z)+{\cal P}(\alpha) \right]_{7C},  
\end{eqnarray} 
where $N_{3C}$, $N_{6C}$ and $N_{7C}$ are, respectively, 
the number of sources detected in each of the simulated 
surveys 
with a particular parameter set used in the model. 
As noted above, if the simulations ``detect'' too many sources as compared to the data, 
then each of the resulting simulation survey samples for 3C, 6C, 7C 
are reduced by 
uniformly removing sources to make the final 
simulation sample sizes equal to that of the data samples. 

The second figure of merit we employ 
adds the KS statistic probabilities for $P$ and $z$ to twice the 
probability for $D$, i.e. ${\cal P}(P)+2{\cal P}(D)+{\cal P}(z)+{\cal P}(\alpha)$ 
for the three surveys, using the same weighting method. 
We denote this as ${\cal P}_{[P, 2D, z, \alpha]}$. 
This second choice was considered because results for $P$ and $z$ usually 
correlate (due to flux-limit arguments); 
thus double weighting the probability for $D$ 
dilutes the impact of the $[{\cal P}(P), {\cal P}(z)]$ correlation. 
Unsurprisingly, in most of the runs we have performed the combined test statistics 
${\cal P}_{[P, D, z, \alpha]}$ and ${\cal P}_{[P, 2D, z, \alpha]}$ 
behaved in a similar fashion. 

Unfortunately, 
for the complicated problem of radio source cosmological evolution,
which involves many parameters and several dimensions, 
any figure of merit based upon 1-D KS tests is a crude approach
in comparing models with observations. 
We attempted to use multi-dimensional statistical tests 
\citep*[e.g.,][]{holmstrom95, loudin03} which, in principle, 
could yield a more robust single figure of merit for the fit 
of our distributions to the data in $\left[P, D, z, \alpha \right]$ space. 
Unfortunately, the limited sizes of the observational samples ($< 150$) 
preclude obtaining reliable results from such generalizations of the KS test. 
Here we are trying to fit four variables, namely 
$\left[P, D, z, \alpha \right]$; in practice the minimum useful sample size 
required would be $\sim 10^4$ for a four-dimensional test. 
In future work we plan to expand our method to include simulations 
of large scale radio surveys containing many thousands of sources, 
such as NVSS and FIRST, which can be made adequately complete in $z$ 
through optical identifications from SDSS 
\citep{ivezic04}. 
Then we might successfully incorporate a multi-dimensional test. 

We call the parameters governing the power evolution as given in the 
papers KDA, BRW (including the parameters used for the initial 
radio source population generation) and MK to be the default ones. 
For KDA and MK 
the authors discuss some alternate parameter sets in the respective papers, 
our defaults are their first and favored parameters. 
To save space, the statistics of only a subset of all 
the runs we have performed are shown in the present work. 
We include the cases which give the best 1-D KS statistical results. 

We present the KS test results in tables grouped by radio source evolution model, 
with each entry illustrating a different parameter set. 
Table 3 gives our results for the KDA model, 
Table 4 for the BRW model, and Table 5 for the MK model. 
The tables for each model follow the same format and pattern. 
Hence we describe only the table entries for the KDA model (Table 3). 


Each of the Tables 3, 4 and 5 give the individual KS statistic 
probabilities ${\cal P}(P)$, ${\cal P}(D)$, ${\cal P}(z)$ 
and ${\cal P}(\alpha)$ for some of the initial runs. 
The results for each model are given in three consecutive rows. 
The first column lists the values of the RG source distribution index $x$ and 
$T_{MaxAge}$ (in Myr) used for the initial population generation in that model run; 
these two parameters were expected to be most important in governing 
the numbers of acceptable sources each Monte Carlo simulation would generate. 
The second column 
first lists the parameter(s) which has (have) been varied from the default case 
in the top row(s), and then gives the initial population (ensemble) size 
used for 3C simulation in that model. 
The third column notes to which survey KS probabilities given 
in the next columns correspond. 
The first row in columns 3, 4, 5, 6 and 7 gives the 3C results, 
with the second and third rows giving the 6C and 7C results, respectively. 
The fourth, fifth, sixth and seventh columns show the values of 
${\cal P}(P)$, ${\cal P}(D)$, ${\cal P}(z)$ and ${\cal P}(\alpha)$ 
respectively for each of the surveys, 3C, 6C and 7C. 
The final, eighth column, lists the combined probabilities, 
${\cal P}_{[P, D, z, \alpha]}$ and ${\cal P}_{[P, 2D, z, \alpha]}$, 
in two consecutive rows, for each particular parameter set. 




To begin with, an initial population, generated using the default parameters 
from BRW for RG population generation, was evolved according to the three 
different default models discussed before. 
The simulated sources detected (according to the prescription in \S4.3) 
were compared to the actual data in the 3C, 6C, and 7C catalogs. 
As shown by the KS test statistics of the first three rows of 
tables 3, 4 and 5, the model fits are all very poor. 
The main problem is that too many high-$z$ and too few low-$z$ sources were 
produced by the models as compared to the data. 


\subsubsection {Dependence on Source Slope Parameter, $x$} 

To look for improved agreement between simulation and data, 
we decided to steepen the beam power distribution 
function of the sources generated in the initial population. 
This significant modification was expected to produce fewer high $P$ -- high $z$ sources, 
and the exponent in the power law distribution of the jet powers, $x$, 
was increased from $x=2.6$ (as used by BRW) in intervals of $\Delta x=0.2$ or $0.3$. 
For the KDA and MK models the overall statistics improved the most at $x=3.0$, 
but were less good for $x=3.3$ or $3.6$. 
For BRW the $P$ and $z$ fits were best for $x=3.6$, 
making the overall performance look fairly good, 
but the $D$ fits were all very bad. 
As will be discussed further below, the former modification ($x=3.0$) 
produced a clear overall improvement for the BRW model too. 


The initial population generated with $x=3$ (but otherwise using the 
BRW prescription), was evolved according to the KDA and MK models. 
The corresponding KS statistics are given as the third entries in Tables 3, 4 and 5. 
For the BRW model, the big population generated using $x=3.6$, 
had a very strong $P-D$ anticorrelation, 
producing too many small sources and too few large ones. 
The combined KS statistics were also much worse than those of the 
KDA and MK models, so we do not list any BRW model results with $x>3.0$. 

Some of the 12 KS probabilities for the KDA and MK models 
(albeit very few for BRW) provide acceptable fits to the data. 
To search for possible further improvements we varied the other parameters 
describing the power evolution in the models as described below. 


Accepting $x=3.0$ as a tentative value for the exponent of the 
beam power distribution for the generated initial population, 
we then varied the parameters 
governing the lobe power evolution of KDA and MK models. 
For BRW the exponent $x=3.6$ was initially accepted as it gave 
good fits for $P$ and $z$, though as noted above, 
the $D$ fit was very poor, and we do not display these results. 
Simulations were done by setting the parameter values at the end points 
of physically reasonable ranges; for example 
we might perform two additional runs using the same initial population 
but we would set a parameter to half or twice its default value. 

Simulated surveys were constructed using the parameter listing 
given in Table 3 (each variation done one at a time) for the 
KDA power evolution model. 
Simulations done with higher axial ratios 
($R_T = 2.0, 2.5, 3.0, 4.0, 5.0$) which are favored by morphological 
data, all yielded severe underdetections 
when compared to the actual number of sources in the catalog. 
Hence we adopted the default value of the axial ratio, 
$R_T = 1.3$, as did KDA. 

The results for the changes of parameters considered 
for the MK power evolution model are given in Table 5. 

As seen from the tables, 
several of the 12 KS probabilities for some cases give acceptable fits, 
but it is difficult to find a single model where all are really good fits. 
In other words, none of the models discussed here simultaneously provide good fits 
to the data from all of the three radio surveys considered. 
As noted above, $P$ and $z$ seem to correlate together in most cases 
because they are related 
when we pick up radio sources by imposing a flux limit on them. 
In some cases $P$ and/or $z$ fits are good and those to $D$ are bad; 
and vice versa. 
The fits to $\alpha$ are almost always poor. 
The KS statistics for model runs which gave any further improvement 
over the ``improved'' default case ($x=3$, Default Model Parameters) 
can be found from Tables 3 -- 5. 




\subsubsection {Dependence on RG Maximum Age} 

An important parameter for the generation of the initial population of 
radio sources according to the BRW prescription is $T_{MaxAge}$. 
It defines the mean active lifetime of the RG central engine and how long 
the radio lobes (being fed by jets powered by AGN activity) 
continue to expand. 
Hence it is one of the most important parameters to constrain if 
we are to estimate the fraction of the relevant volume of the 
universe occupied by radio galaxies during the quasar epoch (\S1). 
As our ultimate goal involves this relevant volume fraction, 
we aim to find the value of $T_{MaxAge}$ which gives the best fit 
to the data for each of the RG evolution models. 
We performed simulation runs with default parameters for each of the models, 
using initial populations with 
$x=3$ (which gives the least bad fits); 
we then set $T_{MaxAge}$ to values in the range 50-600 Myr 
(in intervals of 50 Myr), and obtained the following results. 

For the KDA model, the combined KS probabilities, 
${\cal P}_{[P, D, z, \alpha]}$ or ${\cal P}_{[P, 2D, z, \alpha]}$ 
lacked a single maximum over the range in maximum age considered, 
and peaked at both 150 Myr and 500 Myr. 
However the higher peak was adopted, and hence the adopted best 
maximum age is $T_{MaxAge}=150$ Myr. 
In the other two models, the combined KS probabilities, 
${\cal P}_{[P, D, z, \alpha]}$ or ${\cal P}_{[P, 2D, z, \alpha]}$, 
varied smoothly over the range in maximum age considered. 
In the BRW model the single peak was at $T_{MaxAge}$ = 250 Myr, 
and in the MK model it was at $T_{MaxAge}$ = 150 Myr; 
hence these were adopted for the subsequent runs. 

Monte Carlo runs were done with the above best $T_{MaxAge}$ for 
each model and with $x=2.6$ (the default from BRW), 
to check if that was better. 
For BRW the best $T_{MaxAge}$ when combined with $x=3.0$, 
produced better statistics (less bad $D$ fit), 
and was hence adopted for later runs. 
In all cases $x=3.0$ was better than $x=2.6$ by a significant margin. 
The supporting KS statistics are given in Tables 3, 4 and 5 
for KDA, BRW and MK, respectively. 
Hence we used initial populations with $x=3.0$ and the above ``optimal'' 
$T_{MaxAge}$ values for each model in subsequent runs. 

During the simulation runs we found that a few (for KDA and MK models) 
and some more (for BRW) very small sources 
($D < 1$ kpc) were being ``detected'' in the three modeled surveys (mostly in 3C). 
The actual survey data has negligible numbers of such small sources 
which would not normally be considered FR II types. 
As any such small source (with $D < 1$ kpc) will not be regarded as a 
FR II radio galaxy, we decided to put a linear size cut-off in our simulations. 
For the KDA and MK models a cut-off of 1 kpc was adopted. 
For the BRW model we found that a cut-off of 10 kpc gave much better fits 
than did a 1 kpc cut-off. So in BRW we considered sources only with total 
linear sizes greater than 10 kpc. 
The KDA and MK simulations did not produce many sources with 
linear size $<10$ kpc, hence it did not make much of a difference 
if we imposed a 10 kpc or a 1 kpc size cut-off. 
In the results presented henceforth, these $D$ cut-offs have been incorporated. 

\subsubsection {Dependences on Other Model Parameters} 




In order to further explore the parameter space of the models 
in search of better fits, initial ensembles were generated 
using $x=3.0$, $T_{MaxAge}=150$ Myr for the KDA and MK, 
and $x=3.0$, $T_{MaxAge}=250$ Myr for the BRW simulations
(see \S5.2.3). 
The following prespription was then followed for each. 
The sources in one large random population were evolved 
several times, 
according to one of the three radio lobe power evolution models. 
During each evolution one of the model parameter values 
was varied around its default value (as in \S5.2.2). 
All the parameters of each model given in Table 1 were varied, 
with only one variation per evolution run. 

Only those parameter variations that gave any improvement in statistics 
over the default parameter case of the same model, 
or were essentially as good as the default, were considered further. 
For these parameter sets 
three more initial populations were generated having the same size 
and the same $x$ and $T_{MaxAge}$ values for the different models, 
but with different pseudo-random seeds.
These additional ensembles were then evolved using the 
``improved'' parameter sets 
for each model, and the 1-D KS statistics were found. 
At this point we had the KS test results for a set of four simulations 
of each of the ``improved'' parameter variations. 

The three cases involving variations of a single parameter 
(previous paragraph) which gave the best statistics 
(highest mean ${\cal P}_{[P, D, z, \alpha]}$ of the 4 runs) were then found. 
Simulations were then performed in which two of those parameter changes 
giving better fits were simultaneously employed. 
If these ``2-change'' variations continued to give better performances, 
all three changes were incorporated together in a single run, 
to see if yet better fits could be obtained. 

\subsubsection {Spectral Index ($\alpha$) Behavior}

The spectral index ($\alpha$) at the rest frame of a source
at 151 MHz was estimated 
for each source in the simulated surveys by considering
$\log$ [$\nu$ (MHz)] as the independent variable
and $\log (P_{\nu})$ as the dependent.
The specific powers at the $T_{obs}$ corresponding
to the source (\S4.3) were calculated
at three frequencies, namely,
$151$, $151/(1+z)$ and $151(1+z)$ MHz.   
A quadratic polynomial
was fitted to the $\log (P_{\nu})$ vs. $\log$ [$\nu$ (MHz)] data.
The fit coefficients $a_1$ and $a_2$ where
$\log P_{\nu} = a_0 + a_1 \log \nu + a_2 (\log \nu)^2$, were obtained.
These were used to find the spectral index as
$\alpha = -a_1 - 2 a_2 \log (151/(1.0+z))$.

The KS tests for the fits for the spectral index ($\alpha$)
for all surveys employing each model 
are uniformly bad (as indicated from the ${\cal P}(\alpha)$ values in Tables 3 -- 5).
The poor qualitative fits to the $\alpha$ distribution were
already noted by \citet{BRW} for their models.
Still, it is the BRW model which gives the least
unsatisfactory KS statistics for $\alpha$ fits.

The KS statistics for $\alpha$ fits were extremely bad for the KDA model.
Here, the spectral index distributions consist of a
dense cluster at $\alpha \sim 0.58$,
with no sources having smaller $\alpha$ while some have
steeper spectral indices up to $\alpha \sim 1.0$.
There is a weak $\alpha - D$ anti-correlation until
$D \sim 10^3$ kpc, after which there is a trend of
increasing $\alpha$ as $D$ increases;
but this involves only a few giant sources.

The BRW model also produced mostly very poor $\alpha$ fits,
but occasionally it gave quasi-acceptable KS statistics,
with ${\cal P}(\alpha) \sim 0.01$.
Here, the spectral indices are almost uniformly distributed
within $\alpha \sim 0.58 - 0.85$,
with some sources at smaller $\alpha$.
There is also a weak $\alpha - D$ anti-correlation in the BRW model
which extends throughout the simulated results.

The MK model produced the worst KS statistics for $\alpha$.
Here, the spectral indices came out very steep,
with $\alpha > 0.9$ almost always found.
The distribution includes a cluster at $\alpha \sim 0.9 - 1.0$,
and an extension to very steep spectra $(\alpha \sim 1.5)$.
Here there is a clear trend for $\alpha$ to be higher as $D$ increases
in the simulations for all three catalogs.

Thus, it is clear that all of the models considered to date require
modifications if they are to produce adequate representations of the
observed radio spectral indices.
Making such modifications is a key goal of our future work.

\subsection{Additional Statistical Tests} 

In order to check the robustness of the 
quantitative tests we performed some additional 
statistical analyses. 
We selected the cases of parameter variations 
that gave the highest combined probability, ${\cal P}_{[P, D, z, \alpha]}$, 
of each model, according to the amplified 1-D KS test results (described in \S5.2.4). 
We compared these nominally superior parameter sets for each model 
with the default versions (those with no parameter changes) 
by performing additional statistical tests on them. 

\subsubsection{2-Dimensional Kolmogorov-Smirnov Tests} 

We used the 2-dimensional (2-D) KS test procedure from 
\citet{press02}, which is based on the work of \citet{fasano87}, 
which is a variant of an earlier idea due to \citet{peacock83}. 
The relevant 2-dimensional 2-sample KS probabilities 
(or the significance level indicating that 
the two populations are drawn from the same distribution), 
${\cal P}$, give a quantitative measure of the model fits. 
The comparisons of the model simulated samples to the real data samples 
are done in a way analogous to that for the 1-D KS tests (\S4.3). 

The 2-D KS probabilities for comparisons of 
the properties $P, D, z$ and $\alpha$, taken two at a time, 
for the data and the models, were computed. 
Table 6 shows 
results for both the default versions and the parameter sets giving the 
highest total 1-D KS probability, 
denoted as {\it varied} in the table. 
The results are listed in a similar way as are the 
1-D KS statistics in previous tables. 
The first column gives the model and parameter variation (if any). 
The third, fourth, fifth, sixth, seventh and eighth columns 
list the KS probabilities for comparisons of 
$[P-z]$, 
$[P-D]$, $[z-D]$, $[P-\alpha]$, $[z-\alpha]$ and $[D-\alpha]$ respectively; 
in each case the three rows give results for 3C, 6C and 7C, respectively. 

It is non-trivial to compare the models as there are  
18 values of ${\cal P}$ which must be considered. 
The general trends are discussed in \S6. 



\subsubsection{Correlation Coefficient Analysis} 

We considered the Spearman partial rank correlation coefficients 
between the four relevant source characteristics $P, D, z$ and $\alpha$. 
Following \citet{macklin82}, 
we calculated the partial rank correlation coefficients with four variables, e.g., 
\begin{equation} 
r_{PD, z\alpha} = \frac{r_{PD, z} - r_{P\alpha, z} r_{D\alpha, z}} 
	{\left[ \left(1-r_{P\alpha, z}^2\right) \left(r_{D\alpha, z}^2\right) \right]^{1/2}}, 
\end{equation} 
for the correlation between $P$ and $D$ independent of $z$ and $\alpha$. 
Here the three-variable partial correlation coefficient is
\begin{equation}
r_{PD, z} = \frac{r_{PD} - r_{Dz} r_{Pz}} {\left(1-r_{Dz}\right)^2 \left(1-r_{Pz}\right)^2},
\end{equation}
with $r_{PD}$ being the Spearman correlation coefficient between two variables $P$ and $D$. 

The significance level associated with the 4-variable correlation is 
\begin{equation}
\Sigma_{PD, z\alpha} = \frac { \left(N_{samp}-5\right)^{1/2} } {2}
                       \ln \left( \frac{1+r_{PD, z\alpha}} {1-r_{PD, z\alpha}} \right),  
\end{equation} 
where, $N_{samp}$ is the size of the sample considered. 

The relevant Spearman partial rank correlation coefficients 
involving $[P, D, z, \alpha]$ for the data and for the models 
for those cases for which the 2-D KS tests were done 
are given in Table 7. 
The four-variable correlation coefficients ($r_{PD, z\alpha}, r_{Pz, D\alpha}$, etc) 
were computed by combining the observed data or the model ``simulated'' data 
for all the relevant surveys: 3C, 6C and 7C III. 
We do so in order to dilute the tight $[P-z]$ correlation 
in a single flux-limited complete sample (BRW), 
and to detect the correlations which exist between other characteristics. 
The 2-variable correlation, $r_{PD}$, was always negative; 
however, in the 4-variable case,  
with the effects of $z$ and $\alpha$ removed, 
$r_{PD, z\alpha}$ showed a small positive correlation.
We also examined the correlation coefficients 
of the data and model simulations in 
each survey, 3C, 6C, or 7C, separately. 
\begin{figure*}
\centerline{\epsfig{file=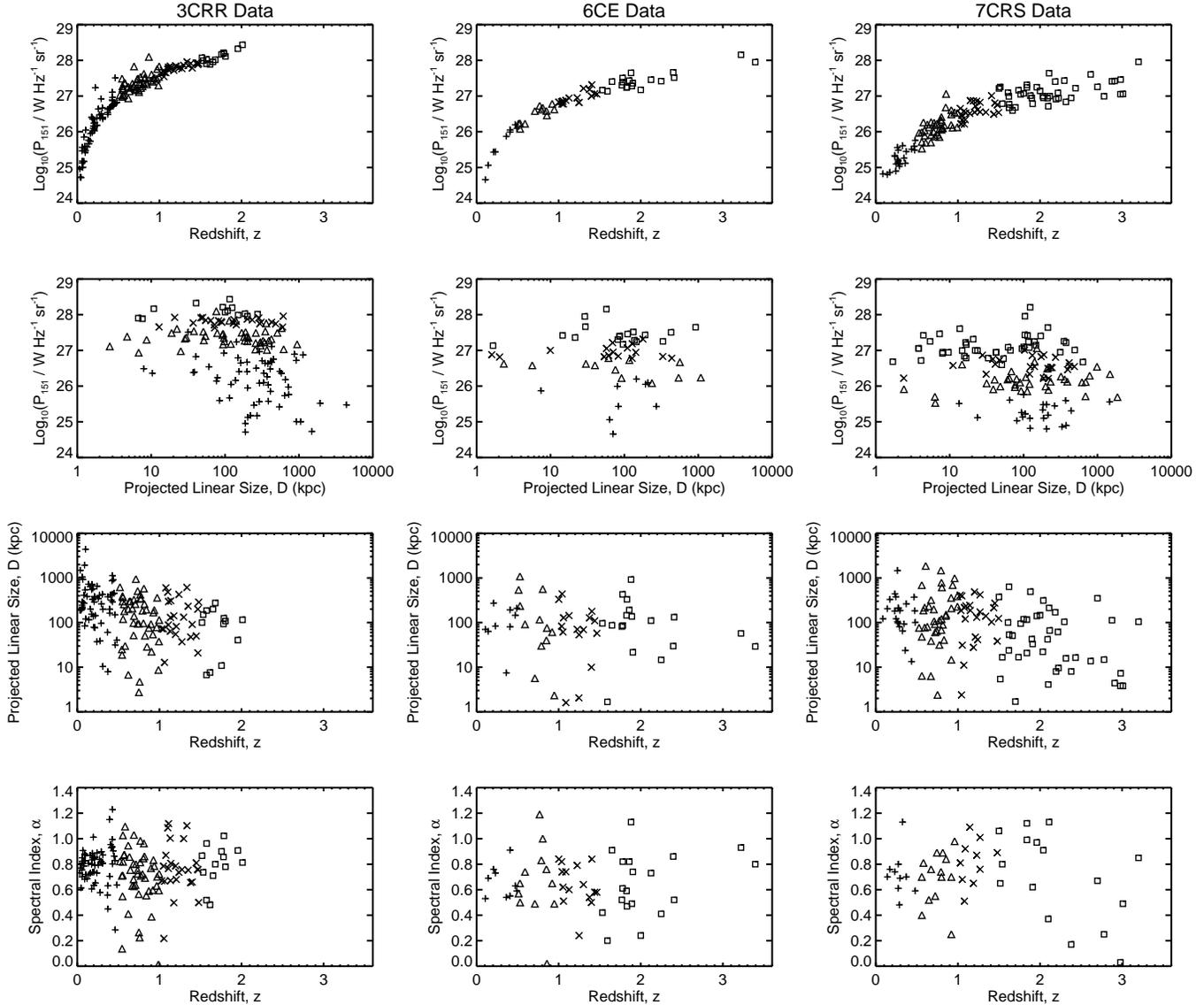, scale=0.9}} 
\caption{The $[P - D - z - \alpha]$ planes for the observational samples 3CRR, 6CE and 7CRS.
The symbols classify the sources into redshift bins as follows;
{\it Plus}: $0 \leq z < 0.5$, {\it Triangle}: $0.5 \leq z < 1.0$,
{\it Cross}: $1.0 \leq z < 1.5$, {\it Square}: $1.5 \leq z$.
The $P-z$ correlations, arising from the flux limits, are clear in the first row.
The $D-z$ plane shows the decreasing trend of average size as redshift increases.
}
\label{fig2}
\end{figure*}

\begin{figure*}
\centerline{\epsfig{file=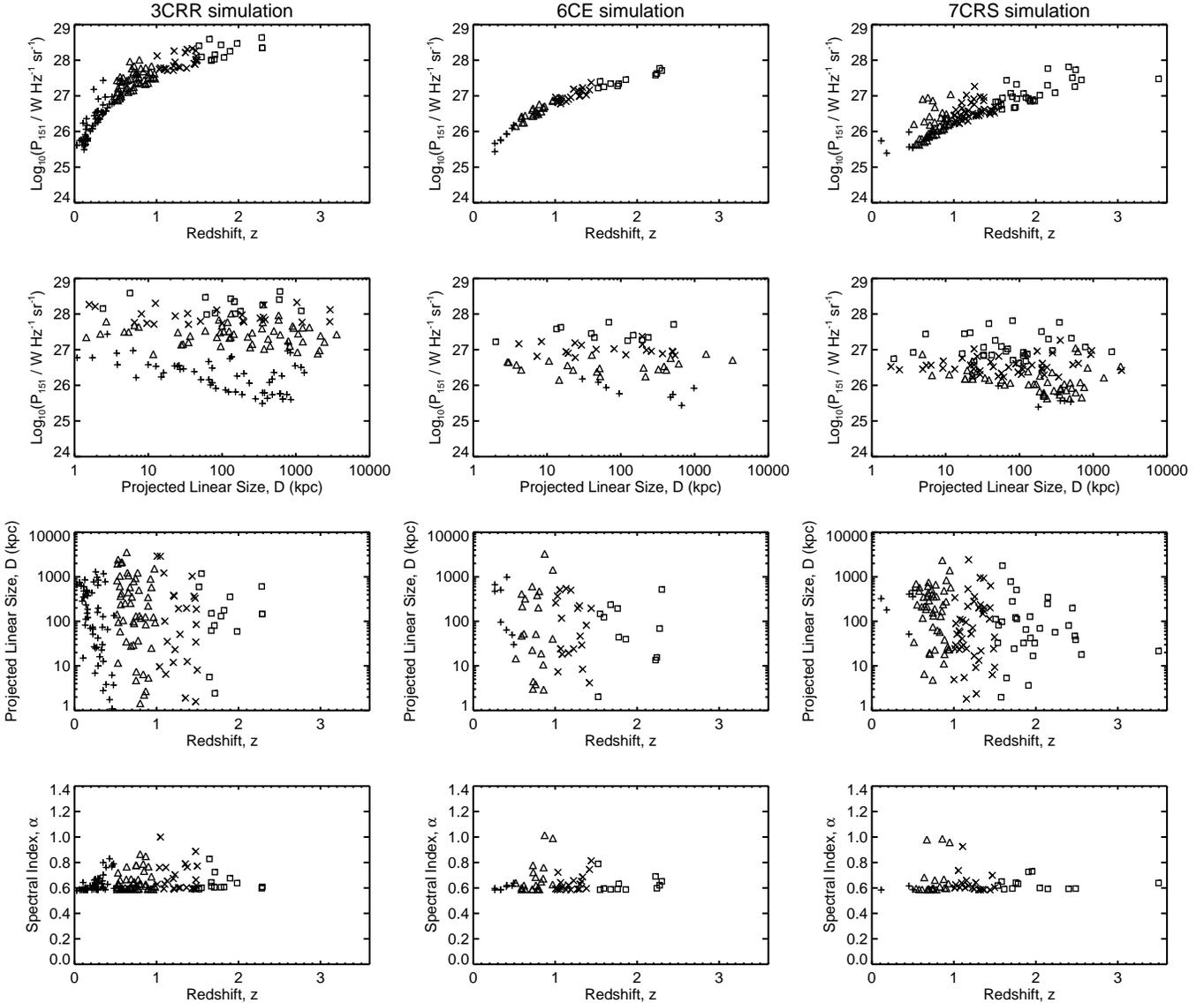, scale=0.9}}
\caption{
The $[P - D - z - \alpha]$ planes for the 3C, 6C and 7C simulations of the KDA Model.
The initial ensemble is formed using $x=3.0$, $T_{MaxAge}=150$ Myr;
the power evolution is with parameter changes
$\rho_0=\rho_{0~({\rm Default})}/2 = 3.6 \times 10^{-22}$ kg m$^{-3}$, $p_m=2.12$,
for a case with initial source population size = 4861474. 
The symbols are as in Fig.~2.
}
\label{fig3}
\end{figure*}

\begin{figure*}
\centerline{\epsfig{file=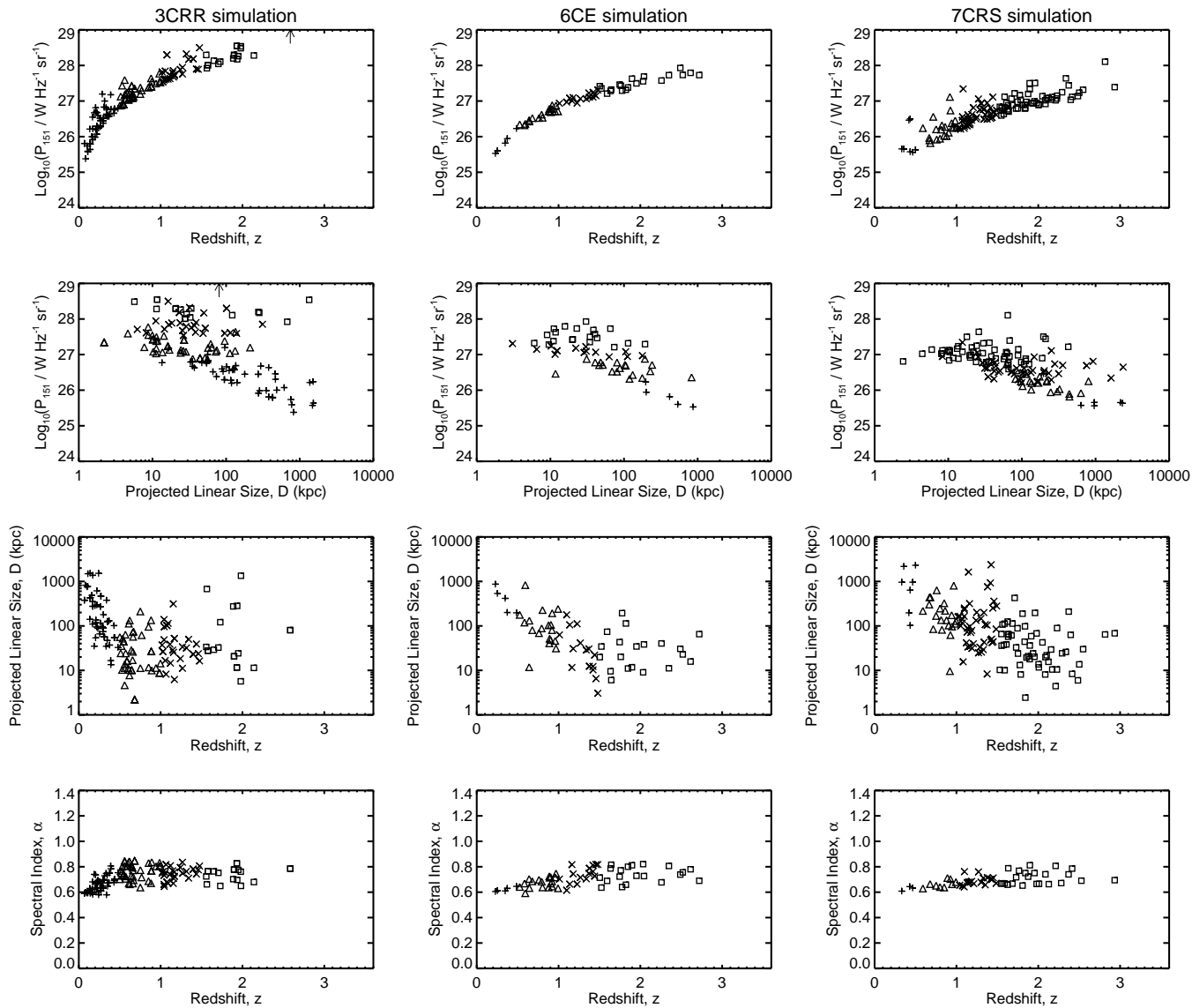, scale=0.9}}
\caption{
The $[P - D - z - \alpha]$ planes for the 3C, 6C and 7C simulations of the BRW Model.
The initial ensemble is formed using $x=3.0$, $T_{MaxAge}=250$ Myr;
the power evolution is with parameter change $a_0=7.5$ kpc,
for a case with initial source population size = 3355926. 
The symbols are as in Fig.~2. 
The upward arrow in the 3C panels implies that one data point exists outside the
plotted range of the figure. 
}
\label{fig4}
\end{figure*}

\begin{figure*}
\centerline{\epsfig{file=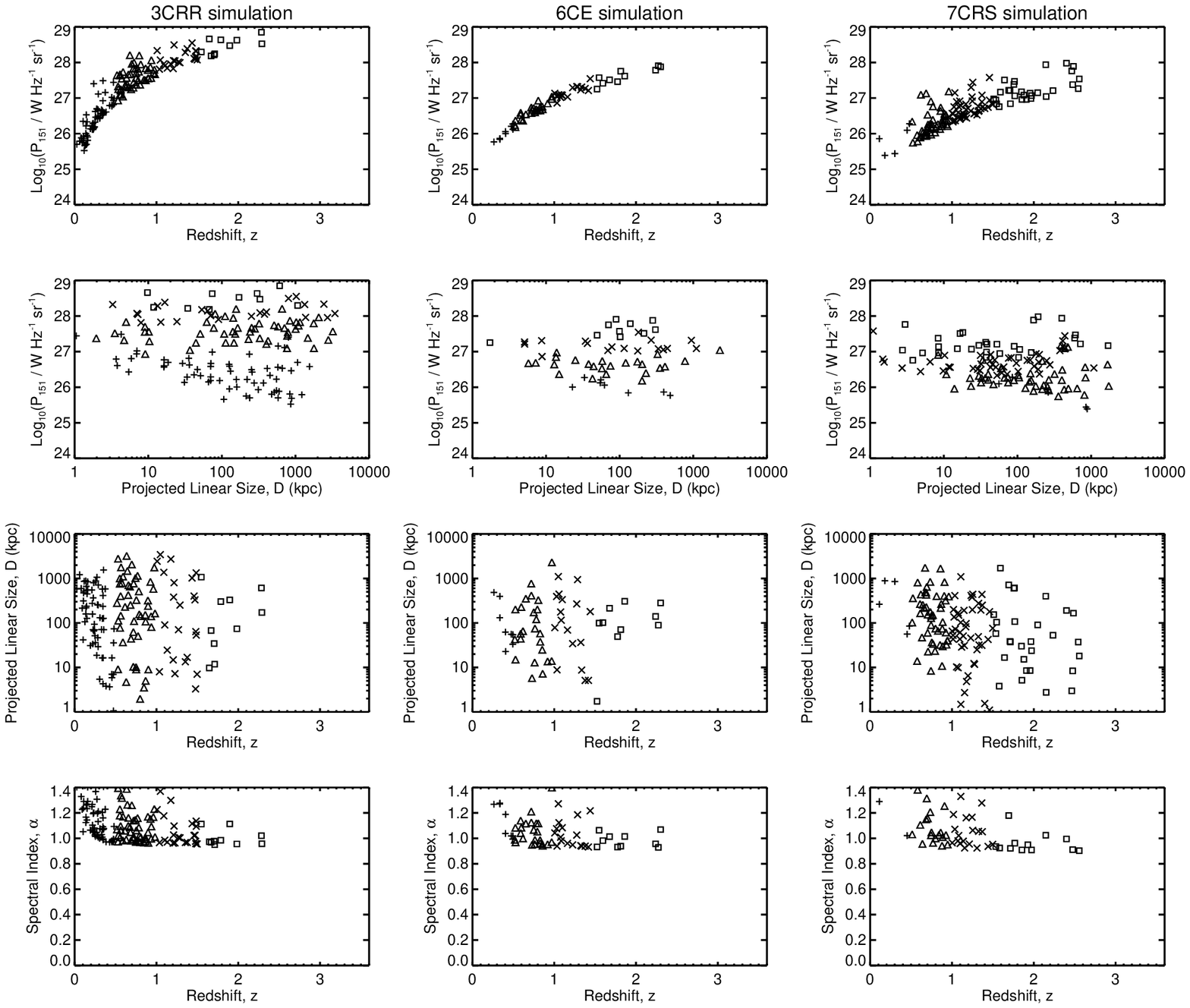, scale=0.9}}
\caption{
The $[P - D - z - \alpha]$ planes for the 3C, 6C and 7C simulations of the MK Model.
The initial ensemble is formed using $x=3.0$, $T_{MaxAge}=150$ Myr;
the power evolution is with parameter change $\gamma_{max(hs)} = 3 \times 10^8$,
for a case with initial source population size = 4861474. 
The symbols are as in Fig.~2.
}
\label{fig5}
\end{figure*}

\section {Discussion} 

We have performed 
quantitative tests of three detailed models for RG evolution. 
This is the first attempt to perform such statistical tests 
involving 4 radio source observables over 3 complete radio surveys. 
During our multi-dimensional Monte Carlo simulation procedure 
we found that it is very difficult to get acceptable simultaneous fits 
to the radio properties $P, D, z$ and $\alpha$ 
for all three redshift complete subsamples of the 3C, 6C and 7C radio catalogs. 
This is true using either the default parameters suggested by 
each of these three leading models, or when considering 
extensive variations upon them involving changing one or 
more of the parameters to plausible different values. 
Usually the $P$ and $z$ fits were correlated, 
due to flux limiting arguments discussed before. 
The fits to the 6C survey were generally better compared to 
those for 3C and 7C; 
this is because of the smaller number of sources in the 
6C catalog and the nature of the KS test. 
Our weighting of the ``total 1-D KS probability'' by the square root 
of the number of sources helps to compensate for this. 
It was most difficult to get acceptable fits to the 
faintest sources cataloged in 7C. 

When varying the model parameters from their default values 
the greatest improvement came from steepening the power law index 
for the initial beam power distribution to $x=3$ 
from $x=2.6$ used by \citet{BRW}. 
This change improved the KDA and MK model performances greatly (Tables 3 and 5). 
The KS statistics for the BRW models were never wonderful, 
especially the $D$ fits (Table 4); 
nonetheless, varying the maximum age assumed for the sources 
from 500 Myr to 150 Myr for the KDA and MK models 
and to 250 Myr for the BRW model also produced better fits. 


We found the following trends for the ratios of number of 
sources detected in the 6C and 7C simulations and the number 
in the actual catalogs, as compared to 3C 
(Ratio$_{6C}$ and Ratio$_{7C}$ respectively). 
For the KDA and BRW models, the detection number ratio was 
more consistent for 6C than for 7C simulations; 
i.e., Ratio$_{6C}$ was closer to 1.0 (which it should equal 
ideally) than was Ratio$_{7C}$. 
For the MK model, the detection number ratios for 6C and 7C 
(which were in the range $0.7 - 1.2$) 
were equally consistent. 
Though we calculated the detection number ratios, 
we do not display them nor did we formally consider them in comparing the models. 
These ratios can be made closer to 1 by varying the 
redshift birth function or the RLF (Eq.~1), 
and so are not good tests of the models per se. 



From the 2-D KS test results we find that 
the $[P-z]$, $[P-D]$ and $[z-D]$ planes can be reasonably fitted
by the ``varied'' models, particularly those for KDA and MK. 
Most of those probabilities are $> 0.2$ for the KDA 
and most exceed 0.05 for MK. 
All of the ${\cal P}$'s of the ``varied'' BRW model not involving $\alpha$ 
are higher than those of the default BRW model.
Improvements are also seen for all of the non-$\alpha$ MK ${\cal P}$'s using the ``varied'' model.
This is the case for only 7 of 9 ${\cal P}$'s of the KDA ``varied'' model. 
These models cannot fit any plane involving $\alpha$, 
with all the $\alpha$-related 2-D KS probabilities $\leq 0.01$ for every model. 
These 2-D results provide support for the hypothesis that the 
``varied'' models based on 1-D KS tests are indeed 
better fits. 

By comparing the values of 2-D KS probabilities in the models of Table 6,
we conclude that KDA model is the best
(having the highest number of ${\cal P}$'s close to 1)
in fitting the observational data,
very closely followed by MK, and finally BRW. 

From the 4-variable correlation coefficient results (Table 7) we see that 
the KDA model is able to match the survey data correlations very closely
(at least for $P, D, z$). 
The matches to the data correlations are less good for the BRW and MK models. 
The parameter variation cases which were the best fits
(i.e., gave highest combined ${\cal P}_{[P, D, z, \alpha]}$)
when judged with respect to 1-D and 2-D KS tests,
are not necessarily the better cases according to the correlation analyses.
The KDA default performs better than the KDA {\it varied} (1-D KS best fit) case.
For BRW and MK models,
the default and the {\it varied} cases perform comparably
(i.e., sometimes the default verison is a better match to the data correlations 
and sometimes the {\it varied} fit is better).

Considering the signs of the four-variable coefficients,
the MK model predicts $[P - \alpha]$ anti-correlation and  $[D - z]$ correlation
which are trends opposite to the survey data and to the other models.
The sign of the $[D -\alpha]$ correlation of the surveys is only predicted by MK,
while the other models produce an anti-correlation; however, 
given the very poor $\alpha$ distribution for the MK model 
this advantage is meaningless. 

From the correlation coefficient analyses we conclude that the 
KDA model fits the data most closely, followed by BRW, and finally MK.
Similar trends are also seen if we examine 
the coefficients obtained by considering each survey separately. 






We plotted slices through the $[P-D-z-\alpha]$ volume 
($P$ vs $z$, $P$ vs $D$, $D$ vs $z$, and $\alpha$ vs $z$) 
for each of the simulated surveys, 
and examined their consistency by comparing them 
with the overall trends in the $[P-D-z-\alpha]$ planes of the actual data. 
The actual data is shown in Fig.\ 2. 
The simulated data are shown in Figs 3, 4 and 5 
for the KDA, BRW and MK models, respectively. 
The plotted simulations are for one of the best 
(in a statistical sense) parameter sets for each model. 
Plots for other good parameter values appear similar, 
while those for worse parameters (according to our KS summary 
statistic) look less like the data. 
Sources are detected out to similar values of redshift, power 
and size in the 3C simulations as in the data. 
The KDA and MK models show very similar trends in $P$, $D$, and $z$. 
The unique features of the BRW model results are discussed below. 

Unsurprisingly, the values of $P$ and $z$ exist in a cluster in the $P-z$ 
plane, above a lower curve determined by the flux limit of the survey. 
All of our simulated surveys of all models miss many of the low $z$ - low $P$ 
sources seen in the data. 
Very high $z$ sources ($z>2.5$) are underproduced in all the 7C simulations, 
and a similar, but less pronounced, trend is also present for 6C. 
All the 3C simulations present a greater scatter in $P$ for 
high $P$ -- high $z$ sources ($z>1$) when compared to the data. 
A few powerful, high $z$ sources are detected in the 3C simulations 
at $z>2.0$ which are not present in the data. 
The scatter in $P$ is naturally less in the 6C survey because of 
the upper (as well as lower) flux limit. 


Examining the $P-D$ planes of the simulations, we find that 
the KDA and MK models overproduce small and large high power sources in 3C, 
and underproduce the large weaker sources. 
The underproduction of low $z$ sources is manifested in the $P-D$ planes 
of the 6C and 7C simulations as the absence of less powerful sources 
(due to the $P-z$ correlation). 


There is a strong $P-D$ evolution seen in the BRW model, 
which is most pronounced in the 3C and 6C simulations. 
The 3C simulation overproduces powerful smaller sources and 
misses several large 
ones. 
The 6C and 7C simulations underproduce less powerful, smaller sources. 
Again, the KDA and MK models show too weak $P-D$ anti-correlations 
than does the data (at least for 3C), 
whereas the BRW model shows too strong an anti-correlation. 

The 6C and 7C simulations show a paucity of low $z$ and high $z$ 
sources in the $D-z$ planes of all the models. 
The KDA and MK models overproduce very small and very large 
3C sources at all redshifts. 
The BRW simulation presents a stronger anti-correlation of 
linear size with redshift, specially for 3C, 
where there are no large sources at intermediate redshifts. 


The $D-z$ evolution (decrease of $D$ as $z$ increases) 
occurs due to imposing survey flux limits. 
This is a ramification of the ``youth -redshift degeneracy'' discussed in \S5.1. 
The high redshift sources show a very steep decline of their 
luminosities with age (seen from the $P-D$ tracks in Fig.\ 1) 
and fall below the survey flux limits at young ages, as their 
radiating particles undergo severe inverse Compton losses off the CMB 
and adiabatic expansion losses as they are transported from 
the high pressure hotspot to the lobes. 
Thus, we can only detect these high $z$ sources at an 
early age when they are still above the limiting survey flux. 
These younger high $z$ sources are naturally smaller and yield the 
weak ``linear size evolution'' (seen in the $D-z$ plane). 
Both the KDA and MK simulations do not show this effect 
as clearly as does the actual data. 
On the other hand, the BRW simulations show stronger 
$D-z$ anti-correlations than do the data.

There are several observational features
(including trends in the $[P-D-z-\alpha]$ planes of the data samples)
that cannot be explained by any models considered so far.
The $[P-D]$ diagram for the 3CRR data show a clear anti-correlation with large scatter.
Another interesting feature is the clump of sources in the 6CE $[P-D]$ diagram near
$D \sim 100$ kpc, $P_{151} \sim 27.5$ W Hz$^{-1}$ sr$^{-1}$ \citep{neeser95}.
Neither of these is reproduced in the models. 
The KDA and MK model simulations predict
too many very large $D > 1$ Mpc and powerful sources (more in 3C, some in 7C),
which are not present in the data. 
This feature has been discussed in \citet{kaiser99a}.

The BRW ${\cal P}(D)$ were very low for many cases (especially for 3C),
and the BRW $[P-D]$ diagrams for all 3 simulated surveys 
showed too strong a $[P-D]$ anti-correlation.
This arises because the BRW model simulations produce too many small but powerful sources.
A possible explanation of this problem could be synchrotron self-absorption
of the radiation emitted by such small powerful sources, 
which is not included in the model. 
Thus some small sources should fall below the survey flux limit 
at a frequency of $151$ MHz. 
Including this effect could improve the relative performance of the BRW model. 

An important point to remember is that
all three models considered here are incomplete in the sense that
they do not incorporate enough physics to predict 
the complete physical conditions prevailing in FR II radio sources. 

Consideration of additional factors may be necessary 
in these models. 
First, the environmental density ($\rho$) could vary with redshift and 
it must eventually deviate from its power law behavior with distance.  
The beam power $(Q_0)$ distribution might vary with redshift 
and the maximum lifetime of AGN activity ($T_{MaxAge}$) 
could vary with redshift and jet power.
Also, the birth function of radio sources with redshift 
(RLF), could have a greater variation with luminosity. 

\section {Conclusions and Future Work} 

We have compared the leading models of radio lobe power evolution for FR II RGs, 
namely the KDA, BRW and MK models, using a simulated radio survey prescription 
(following BRW). 
Each of the dozens of simulated radio surveys we computed required 
the generation and analysis of a few $10^6$ to $> 10^7$ radio sources 
and hence substantial amounts of computing power. 
The total number of Monte Carlo simulations done 
exceeded $250$ and over a billion individual RGs were evolved; 
this was necessary to narrow down the set of parameters 
for each model to the ``best fit'' ranges described in the present work. 
One-dimensional KS tests were done to narrow down the parameters of the 
different models to locate more desirable ones. 
These preferred parameter sets of the models were then compared with the data 
by using 2-D KS 
tests, and correlation coefficient analyses. 

Hydrodynamical modeling of classical double radio sources 
\citep*[e.g.,][]{hooda94, carvalho02} shows that 
the pressure in the nearly self-similarly growing lobes falls with time 
while the hotspot pressure does not vary much. 
The \citet{KDA} model examined here assumed that the head pressure 
falls with time (and is proportional to that of the cocoon), 
so this is a weakness of that model. 
\citet{BRW} adopted a constant hotspot pressure 
(implying more adiabatic losses for particles in the hotspot of older sources) 
while considering the adiabatic expansion of particles 
out of the hotspots to the lobes. 
They showed a rough qualitative agreement between their simulated 
and real 3C and 7C data in the $[P-D-z]$ space. 
\citet{MK} modified the BRW picture by proposing an acceleration mechanism 
occurring throughout the head region; they obtained $[P-D]$ tracks in somewhat 
better accord with 3CRR data, but did not consider $[P-D-z-\alpha]$ distributions. 

Our much more extensive simulations and statistical analyses, 
based on KS tests and correlation coefficients, 
provides a quantitative way to directly compare these three models. 
We note that despite the hundreds of simulations we computed which did employ 
substantial variations on the default  parameters for each RG model 
(only a portion of which are displayed in this paper) we could not completely 
cover the entire plausible parameter space.  We also note that other 
figures of merit could have been devised to distinguish between the goodness of 
fits of the data to the various simulation results, since no really suitable 
multi-parameter statistic is available for samples of this size. 
Keeping these caveats in mind, we believe both that we have covered the 
vast majority of the sensible parameter ranges and that our choice of combined 
KS probabilities is a good way to compare different simulations. 
In this spirit, we now present our conclusions comparing how the models performed 
in different aspects of consistency between the simulations and data.

Our key result is somewhat disappointing. 
Despite investigating a wide range of parameters we find that 
no existing model gives excellent fits to all the data simultaneously. 
However, from the statistical test results, 
the KDA model appears to give better fits than do the BRW or MK models. 

Explicitly judging from the 1-D KS test results, 
the MK model frequently produces acceptable statistics for $P$, $z$ and $D$. 
The KDA simulations also often give adequate statistics. 
The BRW simulations do not give as good statistics as do the MK and KDA models. 
After incorporating the 10 kpc linear size cut-off the statistics for 
some BRW models improve, but are still not as good as 
those given by the other two models. 

According ot the 2-D KS test results, 
the KDA model fits the data most closely, then comes MK, and finally BRW. 
From both the 1-D and 2-D KS test results, 
planes in the $[P-z-D]$ space can be reasonably fitted by 
some parameter variation of the models 
(fits determined by higher values of the KS probabilities), 
but it is difficult to get acceptable $\alpha$ fits. 
Both the KS tests dictate that 
the ``varied'' model parameters of all the models (Tables 6 and 7)
are better fits to the data as compared to the default parameter values. 

However, in terms of reproducing the correlations 
between the source properties (Spearman partial rank correlation coefficient), 
the default models perform better than (KDA) or 
comparable to (BRW, MK) the ``varied'' models. 
The KDA model correlations match the survey data correlations 
most closely, followed by BRW and then MK.  

Our analyses used the redshift birth function of radio sources 
from \citet{willott01}'s radio luminosity function. 
We conclude that, using \citet{willott01}'s RLF, 
the KDA and MK models perform better than BRW in fitting the 
3CRR, 6CE and 7CRS survey data when compared with respect to 
KS-based statistical tests, 
and the KDA model provides the best fits to the correlation coefficients. 

This is the first in a series of papers which aim at comparing the 
performances of radio source evolution models. 
Our goal is to develop one which is a good fit to 
all the observational data. 
We are performing similar tests on a modified model we are developing, 
whose results will be presented elsewhere. 
This new model incorporates conical jet expansion for a fraction 
of a radio source's lifetime within the BRW and MK models. 
This allows us to incorporate a variable hotspot size in the models, 
which is supported by observations \citep{jeyakumar00}. 
Here the hotspot pressure varies as a function of the linear size 
or the source age, which is a more practical possibility 
for hotspots of RGs evolving over 100s of Myr. 



In the future, 
we plan to extend this work by allowing redshift variations in the 
environmental density profile 
(in particular, we will allow for variations of 
$\rho_0$, $a_0$ and $\beta$ with cosmic epoch). 
We also will consider jet propagation through ambient media 
which change from power law density decline to constant densities 
(which change with $z$) at scales around 100 kpc. 
\citet{barai04} gives preliminary work on the implications of 
the volumes attained by radio sources considering cosmological evolution 
of the ambient gas density. 

Our final aim is to estimate the volume fraction of the relevant 
universe occupied by radio lobes, 
and hence to test the robustness of the exciting, but preliminary, 
conclusion that expanding radio galaxies play a significant role 
in the cosmological history of the universe. 

\section*{Acknowledgments} 

We thank the referee, Steve Rawlings, for several useful suggestions 
which substantially improved this paper. 
We thank Katherine Blundell for a helpful conversation, 
Christian Kaiser for conversations and clarifying correspondence, 
Konstantina Manolakou for correspondence and providing us with a version of her Fortran code 
and Chris Willott for correspondence and for sending us the 6C and 7C-III data. 
We are most grateful to Angela Osterman for her efforts on initial versions of some codes, 
and acknowledge conversations with Gopal-Krishna and Zeljko Ivezi{\'c}. 
We also thank Jim Loudin and Hannu Miettinen for correspondence and 
for a version of their multivariate statistics code. 
PJW is most grateful for continuing hospitality at the 
Department of Astrophysical Sciences at Princeton University. 
This work was supported in part by a subcontract to GSU from 
NSF grant AST-0507529 to the University of Washington and by 
Research Program Enhancement funds awarded to the Program in Extragalactic Astronomy at GSU.



\clearpage 

\begin{deluxetable}{ccccccccc} 
\tablewidth{0pc} 
\tablecaption{KDA Model: 1-D KS Statistics for Selected Parameter Variations\label{tab3}\tablenotemark{a}} 
\tablehead{ 
\colhead{$x$} & 
\colhead{Model} & & & & &  &   
\colhead{${\cal P}_{[P, D, z, \alpha]}$ } 
\\ 
\colhead{$T_{MaxAge}$\tablenotemark{b}} & 
\colhead{Ensemble Size} &   
\colhead{Survey} & 
\colhead{${\cal P}(P)$} & 
\colhead{${\cal P}(D)$} & 
\colhead{${\cal P}(z)$} &  
\colhead{${\cal P}(\alpha)$} & 
\colhead{${\cal P}_{[P, 2D, z, \alpha]}$} 
} 

\startdata 
2.6 & Default \tablenotemark{c} & 3C & 9.56e-08 & 0.00366 & 7.65e-07 & 5.76e-08 & 0.899 \\ 
500 & 4397469                   & 6C & 0.537    & 0.00387 & 0.729    & 3.66e-10 & 0.960 \\ 
    &                           & 7C & 0.0254   & 0.0430  & 0.0159   & 0.0295   & \\ 
\\  
2.6 & Default & 3C & 9.07e-12 & 0.00379 & 1.38e-08 & 5.59e-14 & 0.602 \\
150 & 1553389 & 6C & 0.123    & 0.174   & 0.420    & 3.66e-10 & 0.848 \\
    &         & 7C & 7.20e-04 & 0.0834  & 0.00287  & 0.0173   &  \\
\\
3.0 & Default  & 3C & 0.122  & 9.76e-04 & 0.524  & 1.36e-08 & 1.78 \\ 
500 & 11236430 & 6C & 0.819  & 0.0428   & 0.308  & 3.66e-10 & 2.17 \\ 
    &          & 7C & 0.0234 & 0.115    & 0.0159 & 0.0198   &      \\ 
\\ 
3.0 & BRW Env. \tablenotemark{d} & 3C & 0.143  & 1.18e-10 & 0.130  & 2.83e-08 & 0.729 \\ 
500 & $\beta, a_0, \rho_0$       & 6C & 0.458  & 6.85e-07 & 0.129  & 5.33e-09 & 0.730 \\ 
    & 4886474                    & 7C & 0.0367 & 1.09e-04 & 0.0243 & 0.0356   &       \\ 
\\ 
3.0 & $\beta=1.0$ & 3C & 0.00206  & 1.48e-09 & 0.0108   & 0        & 0.0880 \\ 
500 & 485979      & 6C & 0.0325   & 1.10e-07 & 0.0857   & 1.80e-24 & 0.0882 \\ 
    &             & 7C & 2.38e-08 & 7.28e-07 & 8.68e-05 & 3.94e-16 &       \\ 
\\ 
3.0 & $\beta=2.02$ & 3C & 0.207  & 0.0523 & 0.459  & 1.31e-10 & 2.15 \\ 
500 & 9772948      & 6C & 0.979  & 0.509  & 0.529  & 2.20e-09 & 2.60 \\ 
    &              & 7C & 0.0732 & 0.0318 & 0.129  & 0.00211  &    \\ 
\\ 
3.0 & $a_0=1.5$ kpc & 3C & 0.174  & 0.00880 & 0.381 & 1.04e-14 & 2.05 \\ 
500 & 9772948       & 6C & 0.952  & 0.0286  & 0.344 & 3.86e-09 & 2.37 \\ 
    &               & 7C & 0.0524 & 0.154   & 0.371 & 1.88e-04 &     \\ 
\\ 
3.0 & $\rho_0=\rho_1$ \tablenotemark{e} & 3C & 0.0681 & 0.0175 & 0.351 & 1.03e-15 & 2.17 \\ 
500 & 12703438                          & 6C & 0.641  & 0.452  & 0.792 & 6.97e-09 & 2.72 \\ 
    &                                   & 7C & 0.137  & 0.121  & 0.293 & 1.53e-06 &    \\ 
\\ 
3.0 & $\rho_0=\rho_2$ \tablenotemark{f} & 3C & 0.229    & 0.00159 & 0.288   & 1.71e-08 & 1.75 \\ 
500 & 4886474                           & 6C & 0.740    & 0.174   & 0.132   & 3.03e-17 & 2.42 \\ 
    &                                   & 7C & 7.70e-05 & 0.525   & 0.0108  & 6.07e-07 &      \\ 
\\ 
3.0 & $\Gamma_B=5/3$ & 3C & 0.264  & 9.70e-04 & 0.622  & 4.71e-17 & 2.07 \\ 
500 & 7816964        & 6C & 0.952  & 0.0394   & 0.581  & 4.68e-09 & 2.23 \\ 
    &                & 7C & 0.0371 & 0.0867   & 0.0490 & 3.04e-05 &      \\ 
\\ 
3.0 & $\Gamma_B=5/3$ & 3C & 0.0462 & 0.0180 & 0.0104 & 1.91e-15 & 1.51 \\ 
500 & $\Gamma_C=5/3$ & 6C & 0.307  & 0.676  & 0.542  & 1.08e-07 & 2.12 \\ 
    & 14659422       & 7C & 0.129  & 0.162  & 0.0970 & 8.13e-05 &      \\ 
\\ 
3.0 & $R_T=2.0$ & 3C & 0.0913 & 0.00624 & 0.189  & 6.31e-16 & 0.555 \\ 
500 & 11236430  & 6C & 0.269  & 0.355   & 0.355  & 0.0778   & 0.655 \\ 
    &           & 7C & 0.0685 & 0.0283  & 0.0108 & 4.21e-04 &       \\ 
\\ 
3.0 & $\gamma_{min(hs)} = 10$ & 3C & 0.0752 & 3.74e-4 & 0.680  & 8.12e-13 & 2.04 \\ 
500 & 4886474                 & 6C & 0.915  & 0.0683  & 0.413  & 3.66e-10 & 2.43 \\ 
    &                         & 7C & 0.0368 & 0.522   & 0.0689 & 0.0161   & \\ 
\\ 
3.0 & $p$=2.3  & 3C & 0.161 & 0.00370 & 0.282 & 1.44e-07 & 1.65 \\ 
500 & 12703438 & 6C & 0.293 & 0.416   & 0.648 & 9.41e-13 & 1.99 \\ 
    &          & 7C & 0.100 & 0.0457  & 0.237 & 5.45e-06 &      \\ 
\\ 
3.0 & Default & 3C & 0.122  & 6.06e-04 & 0.130 & 4.55e-12 & 1.15 \\
600 & 6615831 & 6C & 0.527  & 0.0680   & 0.182 & 3.66e-10 & 1.38 \\
    &         & 7C & 0.101  & 0.264    & 0.137 & 0.00214  &  \\
\\
3.0 & Default & 3C & 0.320   & 0.274  & 0.360   & 4.74e-12 & 1.84 \\ 
150 & 2652842 & 6C & 0.434   & 0.0141 & 0.585   & 3.66e-10 & 2.34 \\ 
    &         & 7C & 0.00310 & 0.511  & 0.0107  & 2.22e-04 &    \\ 
\\ 
3.0 & $p = 2.12$ & 3C & 0.0911  & 0.0685 & 0.581  & 1.08e-11 & 2.01 \\ 
150 & 4861474    & 6C & 0.444   & 0.164  & 0.413  & 3.66e-10 & 2.79 \\ 
    &            & 7C & 0.00815 & 0.992  & 0.0120 & 2.63e-04 \\  
\\ 
3.0 & $\rho_0=\rho_1$ \tablenotemark{e} & 3C & 0.265  & 0.0186 & 0.515  & 4.89e-17 & 2.16 \\ 
150 & 4861474                           & 6C & 0.879  & 0.244  & 0.308  & 1.29e-09 & 2.72 \\ 
    &                                   & 7C & 0.0271 & 0.626  & 0.0544 & 3.39e-08 \\ 
\\ 
3.0 & $\rho_0=\rho_1$ \tablenotemark{e} & 3C & 0.413 & 0.0502 & 0.766  & 7.42e-21 & 2.33 \\ 
150 & $p = 2.12$                        & 6C & 0.879 & 0.113  & 0.293  & 1.29e-09 & 2.60 \\ 
    & 4861474                           & 7C & 0.105 & 0.232  & 0.0544 & 3.24e-07 \\   

\enddata 
\tablenotetext{a}{These results do not exclude sources with total linear size $D(t) < 1$ kpc.} 
\tablenotetext{b}{$T_{MaxAge}$ in units of Myr.} 
\tablenotetext{c}{Parameter values set equal to those given in the first KDA model \citep{KDA}.} 
\tablenotetext{d}{Parameters defining external environment density profile are set to those of 
                  the BRW model, namely $\beta=1.6, a_0=10$ kpc, $\rho_0=1.67\times10^{-23}$ kg m$^{-3}$.} 
\tablenotetext{e}{$\rho_1 = \rho_{0~({\rm Default})}/2 = 3.6 \times 10^{-22}$ kg m$^{-3}$.} 
\tablenotetext{f}{$\rho_2 = 5 \times \rho_{0~({\rm Default})} = 36 \times 10^{-22}$ kg m$^{-3}$.} 
\end{deluxetable} 
\clearpage 


\begin{deluxetable}{ccccccccc} 
\tablewidth{0pc} 
\tablecaption{BRW Model: 1-D KS Statistics for Selected Parameter Variations\label{tab4}\tablenotemark{a}} 
\tablehead{ 
\colhead{$x$} & 
\colhead{Model} & & & & &  & 
\colhead{${\cal P}_{[P, D, z, \alpha]}$ } 
\\ 
\colhead{$T_{MaxAge}$\tablenotemark{b}} & 
\colhead{Ensemble Size} & 
\colhead{Survey} & 
\colhead{${\cal P}(P)$} & 
\colhead{${\cal P}(D)$} & 
\colhead{${\cal P}(z)$} & 
\colhead{${\cal P}(\alpha)$} &
\colhead{${\cal P}_{[P, 2D, z, \alpha]}$} 
}

\startdata  

2.6 & Default \tablenotemark{c} & 3C & 4.90e-11 & 1.24e-10 & 1.43e-08 & 0.0322   & 0.0564 \\ 
500 & 2930490                   & 6C & 1.93e-05 & 0.0240   & 0.00120  & 3.66e-10 & 0.0728 \\ 
    &                           & 7C & 1.94e-09 & 1.20e-04 & 1.96e-07 & 0.0115   &       \\ 
\\ 
2.6 & Default & 3C & 1.12e-10 & 7.31e-13 & 1.47e-09 & 0.0450   & 0.191 \\ 
250 & 1466378 & 6C & 0.00415  & 9.92e-04 & 0.189    & 3.66e-10 & 0.203 \\ 
    &         & 7C & 5.03e-08 & 0.00101  & 9.96e-09 & 0.0198   &       \\ 
\\ 
3.0 & Default & 3C & 5.59e-08 & 1.99e-19 & 1.55e-04 & 8.59e-05 & 0.0233 \\ 
500 & 2930490 & 6C & 5.06e-04 & 9.50e-05 & 0.0162   & 3.66e-10 & 0.0241 \\
    &         & 7C & 4.17e-07 & 3.70e-05 & 8.07e-07 & 0.0194   &        \\
\\ 
3.0 & Default & 3C & 8.84e-05 & 1.90e-19 & 0.00112  & 1.45e-04 & 0.208 \\ 
250 & 1571349 & 6C & 0.00789  & 2.13e-04 & 0.277    & 3.66e-10 & 0.229 \\ 
    &         & 7C & 1.66e-06 & 1.22e-04 & 2.18e-05 & 0.0104   &      \\ 
\\ 
3.0 & $\beta = 1$ & 3C & 0.00482  & 0.00649 & 0.0448   & 1.44e-07 & 0.507 \\ 
250 & 1571349     & 6C & 0.0893   & 0.113   & 0.214    & 3.03e-17 & 0.772 \\
    &             & 7C & 2.75e-05 & 0.306   & 1.56e-04 & 1.13e-06 \\ 
\\
3.0 & $a_0 = 20$ kpc & 3C & 1.37e-04 & 4.28e-12 & 0.00186  & 1.05e-04 & 0.361 \\ 
250 & 1571349        & 6C & 0.0283   & 0.00415  & 0.214    & 8.69e-14 & 0.566 \\
    &                & 7C & 1.01e-06 & 0.333    & 7.58e-06 & 2.26e-05 \\ 
\\ 
3.0 & $\rho_0=\rho_1$ \tablenotemark{d} & 3C & 1.72e-05 & 1.16e-16 & 0.00287  & 1.02e-07 & 0.229 \\ 
250 & 1571349                           & 6C & 0.0432   & 0.00756  & 0.214    & 1.59e-12 & 0.292 \\
    &                                   & 7C & 6.62e-08 & 0.0957   & 2.08e-06 & 6.91e-04 \\ 
\\
3.0 & $t_{bf} = 100$ yr & 3C & 5.31e-06 & 6.04e-19 & 0.00112  & 4.99e-05 & 0.185 \\ 
250 & 1571349           & 6C & 0.00405  & 4.62e-04 & 0.277    & 3.66e-10 & 0.192 \\
    &                   & 7C & 4.25e-06 & 0.0103   & 7.36e-06 & 0.0237 \\ 
\\  
3.0 & $p = 2.001$ & 3C & 0.264    & 5.03e-06 & 0.680    & 2.87e-04 & 1.39 \\ 
250 & 3355926     & 6C & 0.0473   & 0.511    & 0.141    & 3.66e-10 & 1.72 \\ 
    &             & 7C & 9.07e-05 & 0.00538  & 8.04e-06 & 0.00332  \\ 
\\ 
3.0 & $t_{bs} = 10^3$ Yr & 3C & 0.240    & 1.10e-07 & 0.474    & 0.00487  & 1.42 \\ 
250 & 3355926            & 6C & 0.0481   & 0.509    & 0.310    & 3.66e-10 & 1.89 \\ 
    &                    & 7C & 7.53e-04 & 0.251    & 1.46e-04 & 0.00332 \\ 
\\ 
3.0 & $a_0 = 7.5$ kpc & 3C & 0.330    & 2.90e-09 & 0.668    & 9.34e-07 & 1.63 \\ 
250 & 3355926         & 6C & 0.268    & 0.00764  & 0.583    & 1.29e-09 & 1.73 \\ 
    &                 & 7C & 1.55e-05 & 0.148    & 5.04e-05 & 7.21e-04 \\ 

\enddata 
\tablenotetext{a}{These results are without any total linear size cut-off.} 
\tablenotetext{b}{$T_{MaxAge}$ in units of Myr.} 
\tablenotetext{c}{All other parameters are as given in the BRW model \citep{BRW}.} 
\tablenotetext{d}{$\rho_1 = 2 \times \rho_{0~({\rm Default})} = 3.34 \times 10^{-23}$ kg m$^{-3}$.} 
\end{deluxetable} 
\clearpage 

\begin{deluxetable}{ccccccccc} 
\tablewidth{0pc} 
\tablecaption{MK Model: 1-D KS Statistics for Selected Parameter Variations\label{tab5}\tablenotemark{a}} 
\tablehead{ 
\colhead{$x$} & 
\colhead{Model} & & & & &  & 
\colhead{${\cal P}_{[P, D, z, \alpha]}$ } 
\\ 
\colhead{$T_{MaxAge}$\tablenotemark{b}} & 
\colhead{Ensemble Size} & 
\colhead{Survey} & 
\colhead{${\cal P}(P)$} & 
\colhead{${\cal P}(D)$} & 
\colhead{${\cal P}(z)$} & 
\colhead{${\cal P}(\alpha)$} & 
\colhead{${\cal P}_{[P, 2D, z, \alpha]}$} 
}

\startdata 

2.6 & Default\tablenotemark{c} & 3C & 3.77e-12 & 2.68e-05 & 3.99e-07 & 0        & 0.270 \\ 
500 & 4397469                  & 6C & 0.0283   & 0.0685   & 0.308    & 1.80e-24 & 0.324 \\ 
    &                          & 7C & 5.38e-04 & 0.0119   & 0.0199   & 2.67e-18 &       \\ 
\\ 
2.6 & Default & 3C & 4.05e-14 & 0.00375 & 4.97e-06 & 0        & 0.670 \\ 
150 & 3888492 & 6C & 0.0149   & 0.254   & 0.585    & 1.80e-24 & 0.961 \\
    &         & 7C & 9.77e-09 & 0.413   & 1.18e-05 & 2.16e-15 &       \\
\\ 
3.0 & Default  & 3C & 0.00819  & 0.0356 & 0.229   & 0        & 1.23 \\ 
500 & 11236430 & 6C & 0.0419   & 0.353  & 0.580   & 1.54e-18 & 1.83 \\ 
    &          & 7C & 6.56e-04 & 0.541  & 0.00708 & 2.16e-15 &     \\ 
\\ 
3.0 & KDA Env.\tablenotemark{d} & 3C & 0.126   & 4.01e-04 & 0.353  & 0        & 0.842 \\ 
500 & $\beta, a_0, \rho_0$      & 6C & 0.212   & 0.135    & 0.134  & 3.14e-23 & 0.935 \\ 
    & 14659422                  & 7C & 0.00108 & 0.0149   & 0.0689 & 6.90e-17 &      \\ 
\\ 
3.0 & $\beta=1.6$ & 3C & 0.267  & 0.288 & 0.351 & 0        & 1.73 \\ 
500 & 14659422    & 6C & 0.150  & 0.311 & 0.169 & 7.86e-21 & 2.38 \\ 
    &             & 7C & 0.0523 & 0.647 & 0.176 & 2.16e-15 &      \\ 
\\ 
3.0 & $a_0=7.5$ kpc & 3C & 0.282  & 0.0741 & 0.106  & 0        & 1.76 \\ 
500 & 14659422      & 6C & 0.117  & 0.674  & 0.438  & 1.43e-17 & 2.61 \\ 
    &               & 7C & 0.0839 & 0.802  & 0.0882 & 2.65e-17 &      \\  
\\ 
3.0 & $\rho_0=\rho_1$  \tablenotemark{e} & 3C & 0.400  & 0.0913 & 0.130  & 0        & 1.63 \\ 
500 & 14659422                           & 6C & 0.0770 & 0.205  & 0.335  & 1.54e-18 & 2.30 \\ 
    &                                    & 7C & 0.0839 & 0.756  & 0.0882 & 2.65e-17 &      \\ 
\\ 
3.0 & $\gamma_{min(hs)}=7$ & 3C & 0.170  & 0.0961 & 0.147  & 1.12e-44 & 1.44 \\ 
500 & 14659422             & 6C & 0.361  & 0.624  & 0.376  & 3.54e-16 & 2.00 \\ 
    &                      & 7C & 0.0210 & 0.265  & 0.0794 & 7.04e-21 &      \\ 
\\ 
3.0 & $\gamma_{max(hs)}=3\times10^8$ & 3C & 0.290  & 0.0692 & 0.171 & 0        & 1.53 \\ 
500 & 14659422                       & 6C & 0.404  & 0.0540 & 0.321 & 8.73e-23 & 1.98 \\ 
    &                                & 7C & 0.0685 & 0.519  & 0.132 & 2.16e-15 &      \\ 
\\ 
3.0 & $p$=2.001 & 3C & 0.0917  & 3.68e-05 & 0.354  & 2.99e-27 & 1.35 \\ 
500 & 9772948   & 6C & 0.819   & 0.00394  & 0.183  & 2.60e-20 & 1.48 \\ 
    &           & 7C & 0.00750 & 0.00908  & 0.0970 & 1.94e-08 &      \\ 
\\ 
3.0 & $\epsilon$=1.5 & 3C & 0.00323  & 0.211 & 0.702   & 0        & 1.96 \\ 
500 & 11236430       & 6C & 0.0468   & 0.258 & 0.618   & 3.19e-18 & 2.78 \\ 
    &                & 7C & 4.48e-04 & 0.704 & 0.00779 & 2.76e-18 &      \\ 
\\ 
3.0 & Default & 3C & 0.0277   & 0.420 & 0.633  & 0        & 1.95 \\ 
150 & 4861474 & 6C & 0.310    & 0.420 & 0.182  & 3.14e-23 & 2.92 \\ 
    &         & 7C & 6.75e-04 & 0.267 & 0.0247 & 2.16e-15 &     \\ 
\\  
3.0 & $\rho_0=\rho_1$  \tablenotemark{e} & 3C & 0.161   & 0.0186 & 0.766  & 0        & 1.97 \\ 
150 & 4861474                            & 6C & 0.879   & 0.0135 & 0.0790 & 1.80e-24 & 2.39 \\ 
    &                                    & 7C & 0.00807 & 0.638  & 0.0183 & 1.91e-13 \\  
\\     
3.0 & $\beta = 1.6$ & 3C & 0.0473 & 0.0406 & 0.808  & 0        & 2.37 \\ 
150 & 4861474       & 6C & 0.351  & 0.780  & 0.394  & 2.71e-20 & 3.40 \\ 
    &               & 7C & 0.0238 & 0.942  & 0.0795 & 1.58e-15 \\  
\\ 
3.0 & $\gamma_{max(hs)}=3 \times 10^8$ & 3C & 0.0678  & 0.0361 & 0.668  & 0        & 2.47 \\ 
150 & 4861474                          & 6C & 0.740   & 0.809  & 0.199  & 1.80e-24 & 3.58 \\ 
    &                                  & 7C & 0.00815 & 0.917  & 0.0267 & 1.54e-15 \\ 

\enddata 
\tablenotetext{a}{These results do not exclude sources with total linear size $D(t) < 1$ kpc.} 
\tablenotetext{b}{$T_{MaxAge}$ in units of Myr.} 
\tablenotetext{c}{Parameter values set equal to as given in the preferred MK model \citep[case B,][]{MK}.} 
\tablenotetext{d}{Parameters of external medium density profile are set to those of the default KDA model, 
                  namely, $\beta=1.9, a_0=2$ kpc, $\rho_0=7.2\times10^{-22}$ kg m$^{-3}$.} 
\tablenotetext{e}{$\rho_1 = \rho_{0~({\rm Default})}/1.5 = 1.133 \times 10^{-23}$ kg m$^{-3}$.} 
\end{deluxetable} 
\clearpage 


\begin{deluxetable}{ccccccccc}
\tablewidth{0pc}
\tablecaption{2-D KS Test Results for the Three Models}
\tablehead{
\colhead{Model} & & 
\multicolumn{6}{c}{2-D KS Probability, ${\cal P}$(K-S)} &
\\
\colhead{Parameters} &
\colhead{Survey} &
\colhead{${\cal P}(P - z)$} &
\colhead{${\cal P}(P - D)$} & 
\colhead{${\cal P}(z - D)$} &
\colhead{${\cal P}(P - \alpha)$} &
\colhead{${\cal P}(z - \alpha)$} &
\colhead{${\cal P}(D-\alpha)$} } 

\startdata

KDA                       & 3C & 1.05e-06 & 4.27e-09 & 3.99e-07 & 4.79e-09 & 7.55e-08 & 1.45e-09 \\
Default \tablenotemark{a} & 6C & 0.816    & 0.0164   & 0.00741  & 3.38e-05 & 4.71e-05 & 2.17e-04 \\
                          & 7C & 0.0108   & 0.0124   & 0.00876  & 0.0135   & 0.00763  & 0.00192  \\ 
\\
KDA                      & 3C & 0.531   & 0.0258 & 0.129 & 5.70e-15 & 4.86e-14 & 7.88e-23 \\
Varied \tablenotemark{b} & 6C & 0.445   & 0.370  & 0.244 & 9.88e-04 & 9.15e-04 & 0.00458  \\
                         & 7C & 0.00832 & 0.251  & 0.226 & 8.73e-05 & 3.21e-05 & 5.51e-07 \\ 
\\
BRW                       & 3C & 1.47e-08 & 8.20e-10 & 1.06e-08 & 1.22e-08 & 2.05e-08 & 3.90e-14 \\
Default \tablenotemark{a} & 6C & 3.08e-04 & 0.00116  & 0.00944  & 2.81e-09 & 2.81e-08 & 5.87e-05 \\
                          & 7C & 5.89e-08 & 7.40e-07 & 3.15e-06 & 2.04e-04 & 3.09e-05 & 4.30e-06 \\ 
\\
BRW                      & 3C & 0.432    & 1.34e-07 & 5.70e-07 & 3.68e-05 & 3.36e-05 & 2.71e-17 \\
Varied \tablenotemark{c} & 6C & 0.654    & 0.00918  & 0.0205   & 1.19e-05 & 2.92e-05 & 1.81e-05 \\
                         & 7C & 2.30e-04 & 5.89e-04 & 1.60e-04 & 8.75e-04 & 1.29e-04 & 5.06e-05 \\ 
\\
MK                        & 3C & 5.05e-10 & 3.20e-13 & 3.40e-09 & 4.09e-35 & 1.55e-35 & 3.28e-31 \\
Default \tablenotemark{a} & 6C & 0.0843   & 0.0491   & 0.221    & 2.24e-16 & 3.88e-17 & 3.27e-13 \\
                          & 7C & 1.98e-04 & 4.04e-04 & 4.81e-03 & 4.72e-10 & 2.36e-10 & 9.57e-09 \\ 
\\
MK                       & 3C & 0.0431 & 0.0117 & 0.0701 & 2.24e-35 & 2.73e-35 & 8.38e-34 \\
Varied \tablenotemark{d} & 6C & 0.177  & 0.510  & 0.244  & 2.75e-17 & 9.82e-18 & 1.50e-13 \\
                         & 7C & 0.0247 & 0.0688 & 0.0885 & 3.72e-10 & 1.87e-10 & 1.05e-08 \\

\enddata
\tablenotetext{a}{Simulations with the respective model parameters as used by the authors. 
Initial population generated using $x=2.6$, $T_{MaxAge}=500$ Myr.} 

\tablenotetext{b}{KDA model simulation using initial population with $x=3.0$, $T_{MaxAge}=150$ Myr. 
The power evolution is with parameter changes 
$\rho_0=\rho_{0~({\rm Default})}/2 = 3.6 \times 10^{-22}$ kg m$^{-3}$ and $p=2.12$, 
other parameters set to their default values, 
for a case with initial source population size = 4861474 (last parameter variation entry of Table 3).} 

\tablenotetext{c}{BRW model simulation using initial population with $x=3.0$, $T_{MaxAge}=250$ Myr.
The power evolution is with parameter change $a_0=7.5$ kpc, 
other parameters set to their default values, 
for a case with initial source population size = 3355926 (last parameter variation entry of Table 4).} 

\tablenotetext{d}{MK model simulation using initial population with $x=3.0$, $T_{MaxAge}=150$ Myr. 
The power evolution is with parameter change $\gamma_{max(hs)} = 3 \times 10^8$, 
other parameters set to their default values, 
for a case with initial source population size = 4861474 (last parameter variation entry of Table 5).} 
\end{deluxetable} 
\clearpage


\begin{deluxetable}{cccccccc}
\tablewidth{0pc}
\tablecaption{4-variable Spearman Partial Rank Correlation Analysis \tablenotemark{a}} 
\tablehead{ 
& 
\colhead{Data} & 
\multicolumn{6}{c}{Model (combining all surveys \tablenotemark{a} ~)}
\\
& &
\multicolumn{2}{c}{KDA} &
\multicolumn{2}{c}{BRW} &
\multicolumn{2}{c}{MK}
\\
\colhead{Coeff.} &
\colhead{All \tablenotemark{a}}     &
\colhead{Default} &
\colhead{Varied \tablenotemark{b}}  & 
\colhead{Default} & 
\colhead{Varied \tablenotemark{b}}  & 
\colhead{Default} & 
\colhead{Varied \tablenotemark{b}} }

\startdata

$r_{PD, z\alpha}$ \tablenotemark{c}     
                       & 0.0303  & 0.0528  & 0.198   & 0.102   & 0.0944 & 0.358  & 0.196 \\ 
$\Sigma_{PD, z\alpha}$ \tablenotemark{d} 
                       & 0.478   & 0.844   & 3.20    & 1.63    & 1.51   & 5.96   & 3.16 \\ 
\\
$r_{Pz, D\alpha}$      & 0.716   & 0.668   & 0.648   & 0.415   & 0.576  & 0.569  & 0.495 \\
$\Sigma_{Pz, D\alpha}$ & 14.2    & 12.9    & 12.3    & 7.04    & 10.4   & 10.3   & 8.63 \\ 
\\ 
$r_{Dz, P\alpha}$      & -0.268  & -0.274  & -0.206  & -0.106  & -0.234 & 0.303  & 0.433 \\
$\Sigma_{Dz, P\alpha}$ & -4.33   & -4.48   & -3.33   & -1.70   & -3.78  & 4.97   & 7.37 \\
\\ 
$r_{P\alpha, Dz}$      & 0.147   & 0.0456  & 0.318   & 0.329   & 0.428  & -0.167 & -0.103 \\
$\Sigma_{P\alpha, Dz}$ & 2.33    & 0.729   & 5.25    & 5.45    & 7.27   & -2.68  & -1.65 \\ 
\\
$r_{D\alpha, Pz}$      & 0.472   & -0.0287 & -0.640  & -0.890  & -0.881 & 0.922  & 0.888 \\
$\Sigma_{D\alpha, Pz}$ & 8.08    & -0.459  & -12.1   & -22.7   & -22.0  & 25.5   & 22.5 \\ 
\\
$r_{z\alpha, PD}$      & -0.0234 & 0.0970  & -0.0935 & -0.0237 & -0.226 & -0.465 & -0.569 \\
$\Sigma_{z\alpha, PD}$ & -0.369  & 1.55    & -1.50   & -0.379  & -3.65  & -8.01  &  -10.3 \\ 

\enddata 

\tablenotetext{a}{The four observables $P$, $D$, $z$ and $\alpha$ 
for the 3C, 6C and 7C III surveys (whether real or simulated), 
combined together in a single sample.}

\tablenotetext{b}{The particular parameters used are the same 
as those in Table 6 for each of the KDA, BRW and MK {\it varied} models.}  

\tablenotetext{c}{Spearman partial rank correlation coefficient between 
two variables $P$ and $D$, when the other two variables $z$ and $\alpha$ are kept fixed.} 

\tablenotetext{d}{Significance level associated with the correlation between $P$ and $D$, 
independent of $z$ and $\alpha$.}

\end{deluxetable}


\end{document}